\RequirePackage[2020-02-20]{latexrelease}     %%  Required on iMac Pro to prevent error claiming I was missing \begin{document}
\documentclass[twocolumn,astrosymb]{aastex62}

\pdfoutput=1
\PassOptionsToPackage{pdftex}{graphicx}
\usepackage{morefloats}
\usepackage{epstopdf}
\usepackage{amsmath}
\usepackage{amssymb}
\usepackage{latexsym}
\usepackage{multirow}
\usepackage{tensor}
\usepackage{enumitem}
\PassOptionsToPackage{pdfpagelabels}{hyperref}
\hypersetup{linkcolor=red,citecolor=blue,filecolor=cyan,urlcolor=teal}
\received{\today}
\revised{N/A}
\accepted{N/A}
\submitjournal{ApJS}
\shorttitle{WIND SPACECRAFT POTENTIAL}
\shortauthors{Wilson III et al.}
\watermark{Rough Draft}
\setwatermarkfontsize{50pt}
\makeatletter
\renewcommand\@makefnmark{\hbox{\@textsuperscript{\normalfont\color{violet}\@thefnmark}}}
\makeatother
\newcommand{\totalnvdfsall}{8,804,545}      %%  Total # of VDFs examined, All
\newcommand{\totalnvdfsAs}{4,014,873}       %%  Total # of VDFs examined, A
\newcommand{\totalnvdfsBs}{4,213,734}       %%  Total # of VDFs examined, B
\newcommand{\totalnvdfsCo}{127,514}         %%  Total # of VDFs examined, C1
\newcommand{\totalnvdfsCt}{448,424}         %%  Total # of VDFs examined, C2
\newcommand{\totalnvdfsCs}{575,938}         %%  Total # of VDFs examined, Cs
\newcommand{\totalpvdfsAs}{45.60}           %%  Percent of total # of VDFs examined, A
\newcommand{\totalpvdfsBs}{47.86}           %%  Percent of total # of VDFs examined, B
\newcommand{\totalpvdfsCo}{1.45}            %%  Percent of total # of VDFs examined, C1
\newcommand{\totalpvdfsCt}{5.09}            %%  Percent of total # of VDFs examined, C2
\newcommand{\totalpvdfsCs}{6.54}            %%  Percent of total # of VDFs examined, Cs

\makeatletter
 \DeclareRobustCommand\ref{%
    \@ifstar\@refstar\T@ref
  }%
  \DeclareRobustCommand\pageref{%
    \@ifstar\@pagerefstar\T@pageref
  }%
 \makeatother
\begin{document}
%%  style of bibliography
\bibliographystyle{aasjournal}
%%  Turn page anchors back on
%%\hypersetup{pageanchor=true}

%%  TITLE
\title{Spacecraft floating potential measurements for the \emph{Wind} spacecraft}
%%  AUTHORS AND AFFILIATIONS - 2 methods
\correspondingauthor{L.B. Wilson III}
\email{lynn.b.wilsoniii@gmail.com}

\author[0000-0002-4313-1970]{Lynn B. Wilson III}
\affiliation{NASA Goddard Space Flight Center, Heliophysics Science Division, Greenbelt, MD, USA.}

\author[0000-0002-6536-1531]{Chadi S. Salem}
\affiliation{Space Sciences Laboratory, University of California, Berkeley, CA 94720-7450, USA.}

\author[0000-0002-0675-7907]{John W. Bonnell}
\affiliation{Space Sciences Laboratory, University of California, Berkeley, CA 94720-7450, USA.}

%%  ABSTRACT  [≤ 250 words]
\begin{abstract}
  Analysis of \totalnvdfsall~electron velocity distribution functions (VDFs), observed by the \emph{Wind} spacecraft near 1 AU between January 1, 2005 and January 1, 2022, was performed to determine the spacecraft floating potential, $\phi{\scriptstyle_{sc}}$.  \emph{Wind} was designed to be electrostatically clean, which helps keep the magnitude of $\phi{\scriptstyle_{sc}}$ small (i.e., $\sim$5--9 eV for nearly all intervals) and the potential distribution more uniform.  We observed spectral enhancements of $\phi{\scriptstyle_{sc}}$ at frequencies corresponding to the inverse synodic Carrington rotation period with at least three harmonics.  The 2D histogram of $\phi{\scriptstyle_{sc}}$ versus time also shows at least two strong peaks with a potential third, much weaker peak.  These peaks vary in time with the intensity correlated with solar maximum.  Thus, the spectral peaks and histogram peaks are likely due to macroscopic phenomena like coronal mass ejections (solar cycle dependence) and stream interaction regions (Carrington rotation dependence).  The values of $\phi{\scriptstyle_{sc}}$ are summarized herein and the resulting dataset is discussed.
\end{abstract}


%%----------------------------------------------------------------------------------------
%%  Section:  Background and Motivation
%%----------------------------------------------------------------------------------------
\phantomsection   %%  Fix reference link
\section{Background and Motivation}  \label{sec:introduction}

%%  Solar Wind Intro
\indent  The solar wind is a nonequlibrium, kinetic, ionized gas comprised of electrons and multiple ion species (with multiple charge states) from hydrogen up through iron \citep[e.g.,][]{bame68a, bochsler85a, gloeckler98a, lepri13a}.  The solar wind is not in equilibrium because it weakly collisional \citep[e.g.,][]{maruca13b, salem03a}, allowing for the consistent presence of finite heat fluxes and the temperatures of species $s'$ and $s$ are not equal, i.e., $\left(T{\scriptstyle_{s'}}/T{\scriptstyle_{s}}\right){\scriptstyle_{tot}}$ $\neq$ 1, for $s'$ $\neq$ $s$ (see Appendix \ref{app:Definitions} for parameter definitions) \citep[e.g.,][]{wilsoniii18b, wilsoniii19a, wilsoniii19b}.

%%  Spacecraft Charging Intro
\indent  Spacecraft propagating through space experience a net charging, called the spacecraft potential or $\phi{\scriptstyle_{sc}}$ \citep[e.g.,][]{besse80a, garrett81a, geach05a, genot04a, grard73a, grard83a, lai17a, lavraud16a, pedersen95a, pulupa14a, salem01a, scime94a, scudder00a, whipple81a}.  On average, when the spacecraft is in sunlight there is a net-zero current balance between the ambient plasma and the spacecraft.  The current sources are thermal currents from ambient plasma species, $j{\scriptstyle_{th,s}}$, bulk flow velocity currents if in regions like the solar wind, $j{\scriptstyle_{o,s}}$, photoelectron currents due to the photoelectric effect, $j{\scriptstyle_{ph}}$, and currents from emission of secondary particles from impact ionization $j{\scriptstyle_{2nd,s}}$.  The magnitude of $\phi{\scriptstyle_{sc}}$ depends upon the electromagnetic cleanliness of the spacecraft, the space environment, and whether the spacecraft is in sunlight or the shadow of some object.  Generally, however, the better the electromagnetic cleanliness the smaller in magnitude $\phi{\scriptstyle_{sc}}$ will be, on average.  An accurate measure of $\phi{\scriptstyle_{sc}}$ is critical for properly calibrating thermal electron detectors and calculating accurate electron velocity moments \citep[e.g.,][]{geach05a, genot04a, lavraud16a, pulupa14a, salem01a, scime94a}.

%%  Spacecraft Potential Measurement Intro
\indent  Modern spacecraft equipped with electric field probes can directly measure $\phi{\scriptstyle_{sc}}$ versus time, e.g., the Magnetospheric Multiscale (MMS) mission \citep[e.g.,][]{ergun16a, lindqvist16a}.  This requires a constant measurement of the quasi-static, DC-coupled electric potential versus time.  Many earlier spacecraft were much more restricted in onboard memory, telemetry rates, and receiver electronics.  Thus, it is often required that researchers infer the spacecraft potential by relying on discrepancies between two measurements of the same quantity\footnote{Often it is the difference between the total electron density derived from the quasi-thermal noise upper hybrid line and the measured density from a particle instrument.} \citep[e.g., see][]{salem01a, salem23a} or directly measuring the separation between ambient and photoelectrons with something like a thermal electron detector \citep[e.g.,][]{phillips93a}.  The former technique requires both instruments be very well calibrated while the latter technique requires the instrument minimum energy, $E{\scriptstyle_{min}}$, be smaller than $\phi{\scriptstyle_{sc}}$ for all $\phi{\scriptstyle_{sc}}$ $>$ 0 eV.

%%++++++++++++++++++++++++++++++++++++++++++++++++++++++++++++++++++++++++++++++++++++++++
%% Image:  Example Photoelectrons
%%++++++++++++++++++++++++++++++++++++++++++++++++++++++++++++++++++++++++++++++++++++++++
\begin{figure}
  \centering
    {\includegraphics[trim = 0mm 0mm 0mm 0mm, clip, width=80mm]{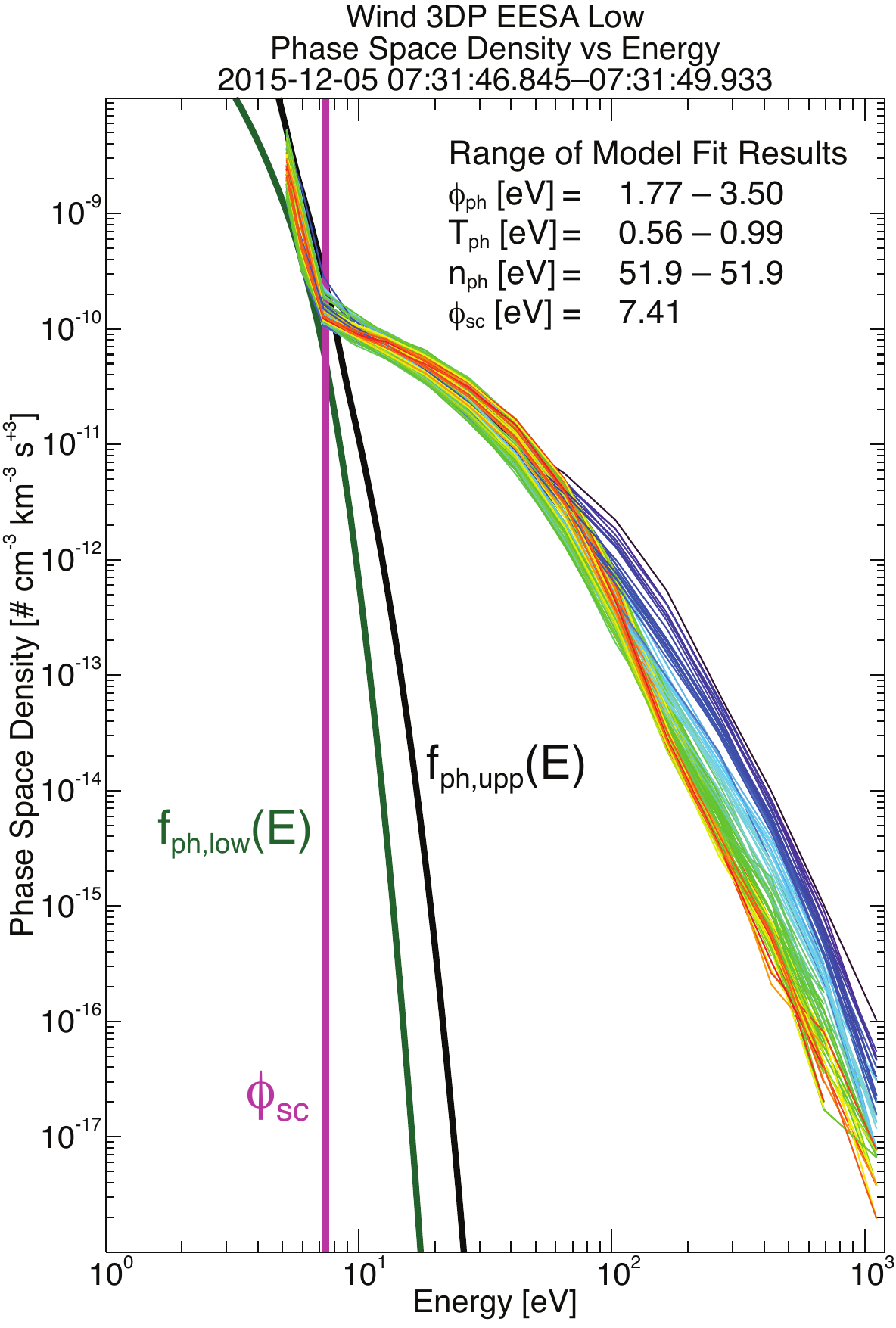}}
    \caption{Example electron energy distribution function (EDF) from the \emph{Wind} 3DP instrument.  The multi-colored thin lines show the phase space density [$\# \ cm^{-3} \ km^{-3} \ s^{+3}$] of each of the 88 solid angle bins versus energy, $E$ [$eV$].  The photoelectrons are seen to the left of the vertical magenta line marking the spacecraft floating potential, $\phi{\scriptstyle_{sc}}$.  The thick green and black lines show simplified model EDFs for the photoelectrons \citep[following method in][]{wilsoniii22h} with parameter ranges shown in the legend of the plot.}
    \label{fig:ExamplePhotoelectrons}
\end{figure}
%%++++++++++++++++++++++++++++++++++++++++++++++++++++++++++++++++++++++++++++++++++++++++
%% Image:  Example Photoelectrons
%%++++++++++++++++++++++++++++++++++++++++++++++++++++++++++++++++++++++++++++++++++++++++

%%  Photoelectron Intro
\indent  For most spacecraft operating in the solar wind, there are few (if any) energy bins that satisfy $E$ $\leq$ $\phi{\scriptstyle_{sc}}$, so the proper shape or form of the velocity distribution function (VDF) for the photoelectrons is not well known \citep[e.g.,][]{phillips93a}.  The simplest approach is to model them as a single, isotropic Maxwellian-Boltzmann distribution \citep[e.g.,][]{grard73a, halekas20a, wilsoniii22h}.  There is some evidence that the photoelectron distribution in the solar wind is not a single Maxwellian but may be at least two populations (\textit{Salem et al.,} in preparation), consistent with magnetospheric observations \citep[e.g.,][]{pedersen95a}.  However, it is beyond the scope of the current study to determine the shape of photoelectron VDF but it is a topic of active investigation.

\indent  This paper is outlined as follows:  Section \ref{sec:DefinitionsDataSets} introduces the datasets and methodology; Section \ref{sec:SolarWindStatistics} presents some preliminary long-term solar wind statistics; and Section \ref{sec:DiscussionandConclusions} discusses the results and interpretations with recommendations for future measurement/instrumentation requirements.  Appendices are also included to provide additional details of the parameter definitions (Appendix \ref{app:Definitions}), some additional one-variable statistics (Appendix \ref{app:ExtraStatistics}), and a detailed description of the public dataset generated from the analysis in this study (Appendix \ref{app:PublicDataset}).

%%----------------------------------------------------------------------------------------
%%  Section:  Data Sets and Methodology
%%----------------------------------------------------------------------------------------
\phantomsection   %%  Fix reference link
\section{Data Sets and Methodology}  \label{sec:DefinitionsDataSets}

%%****************************************************************************************
%%  Subsection:  Data Sets
%%****************************************************************************************
\phantomsection   %%  Fix reference link
\subsection{Data Sets}  \label{subsec:DataSets}

\indent  Nearly all data presented herein were measured by the \emph{Wind} spacecraft \citep[][]{wilsoniii21b} near 1 AU in the solar wind, specifically the thermal electron detector EESA Low, part of the 3DP instrument suite \citep[][]{lin95a}.  The data are taken from both burst and survey modes, which has cadences of $\sim$3 seconds and $\sim$24--78 seconds, respectively\footnote{All products discussed herein start from the \emph{Wind} 3DP level zero files found at \url{http://sprg.ssl.berkeley.edu/wind3dp/data/wi/3dp/lz/}.}.  All EESA Low distributions have spin period durations or total integration times (i.e., $\sim$3 seconds for entire mission).  All VDFs have 15 energy bins and 88 solid angle look directions with typical resolutions of $\Delta \ E$/$E$ $\sim$ 20\% and $\Delta \ \phi$ $\sim$ 5$^{\circ}$--22.5$^{\circ}$ depending on the poloidal anode (i.e., the ecliptic plane bins have higher angular resolution than the zenith).  All in situ measurements presented herein are shown in the spacecraft frame of reference.

\indent  We also present observations of solar emission lines from the Thermosphere Ionosphere Mesosphere Energetics Dynamics (TIMED) spacecraft, specifically the  Solar EUV Experiment (SEE) \citep[][]{woods05a}.  The data are used as a proxy for solar cycle and variations in ionizing photon and photoelectron flux versus time.

\indent  We examined all electron energy distribution functions (EDFs) between January 1, 2005 and January 1, 2022 corresponding to a total of \totalnvdfsall~EDFs.  We chose this time range because the start is well after \emph{Wind} moved to the first Sun-Earth Lagrange point permanently and the detector $E{\scriptstyle_{min}}$ was below $\sim$6 eV.  The lower bound on $E{\scriptstyle_{min}}$ ranged from $\sim$3.2--5.2 eV for all EDFs examined.  It is critical that $E{\scriptstyle_{min}}$ $<$ $\phi{\scriptstyle_{sc}}$ otherwise we can only define an upper bound estimate for $\phi{\scriptstyle_{sc}}$ from the thermal electron measurements.

%%****************************************************************************************
%%  Subsection:  Photoelectron Identification
%%****************************************************************************************
\phantomsection   %%  Fix reference link
\subsection{Photoelectron Identification}  \label{subsec:PhotoelectronIdentification}

\indent  All EDFs were plotted to a computer screen and viewed, ``by eye'', in rapid succession\footnote{The plots were viewed more like a high frame rate movie than a series of individual images to make the effort feasible.} to determine some overarching properties of the electrons.  Several important features were observed from this effort.  The first is that $\phi{\scriptstyle_{sc}}$ $<$ 30 eV for nearly every EDF of the \totalnvdfsall~EDFs examined herein.  Thus, we imposed an a priori maximum energy, $E{\scriptstyle_{max}}$, for any $\phi{\scriptstyle_{sc}}$ solution of 30 eV\footnote{Previous solar wind estimates/measurements of $\phi{\scriptstyle_{sc}}$ in the solar wind typically fall below $\sim$10--15 eV \citep[e.g.,][]{geach05a, pedersen95a, salem01a}.}.  Second, we observed that the EDFs had four characteristic shapes/profiles (discussed in more detail below) that allowed us to simplify our approach.

\indent  To provide the reader with some context, Figure \ref{fig:ExamplePhotoelectrons} shows an example EDF with example model predictions for the photoelectrons.  The electron data are shown as the thin, multi-colored lines while the photoelectron model and $\phi{\scriptstyle_{sc}}$ solution are shown with thick, color-coded lines.  The data are shown in units of phase space density for ease of modeling since these are the natural units of a Maxwell-Boltzmann distribution.  The details of the model are discussed in \citet[][]{wilsoniii22h}.  The model parameters are shown in the figure legend in the upper right-hand corner of the plot (see Appendix \ref{app:Definitions} for symbol definitions).  The purpose of Figure \ref{fig:ExamplePhotoelectrons} is to provide the reader with context as to which parts of the EDF are photoelectrons versus ambient solar wind electrons.

\indent  One may ask how we know the data falling between the thick green and black lines are photoelectrons and not some cold, ambient solar wind population.  We can measure the total electron density, $n{\scriptstyle_{e}}$, from the upper hybrid line \citep[e.g.,][]{meyervernet89a} observed by the \emph{Wind} WAVES instrument \citep[][]{bougeret95a}.  We can invert the frequency of this line, knowing the quasi-static magnetic field, to calculate the electron plasma frequency and thus the electron density.  Doing this gives us $n{\scriptstyle_{e}}$ $\sim$ 7.95 cm$^{-3}$.  For comparison, a typical value for the photoelectron number density, $n{\scriptstyle_{ph}}$, is $>$200 cm$^{-3}$ and the mean kinetic energy of the photoelectrons, or their temperature, $T{\scriptstyle_{ph}}$, is $\sim$1 eV \citep[e.g.,][]{grard73a, grard83a}.

\indent  To further illustrate why the electrons between the thick lines are not part of the ambient solar wind, we can numerically integrate the EDF following the usual approach of adjusting the distribution by $\phi{\scriptstyle_{sc}}$ then calculating the velocity moments \citep[e.g., see][]{wilsoniii19b, wilsoniii22h}.  We find the integrated velocity moments are $n{\scriptstyle_{e, int}}$ $\sim$ 7.01 cm$^{-3}$ and $T{\scriptstyle_{e, int}}$ $\sim$ 22.5 eV.  If we do not remove the photoelectrons and do not adjust the energies prior to integration, we find $n{\scriptstyle_{e, bad}}$ $\sim$ 39.2 cm$^{-3}$ and $T{\scriptstyle_{e, bad}}$ $\sim$ 7.34 eV.  Since the upper hybrid line density is $\sim$7.95 cm$^{-3}$, it is clear that the electrons below $\sim$7.41 eV are photoelectrons.

%%++++++++++++++++++++++++++++++++++++++++++++++++++++++++++++++++++++++++++++++++++++++++
%% Image:  Example EDF
%%++++++++++++++++++++++++++++++++++++++++++++++++++++++++++++++++++++++++++++++++++++++++
\begin{figure}
  \centering
    {\includegraphics[trim = 0mm 0mm 0mm 0mm, clip, width=80mm]{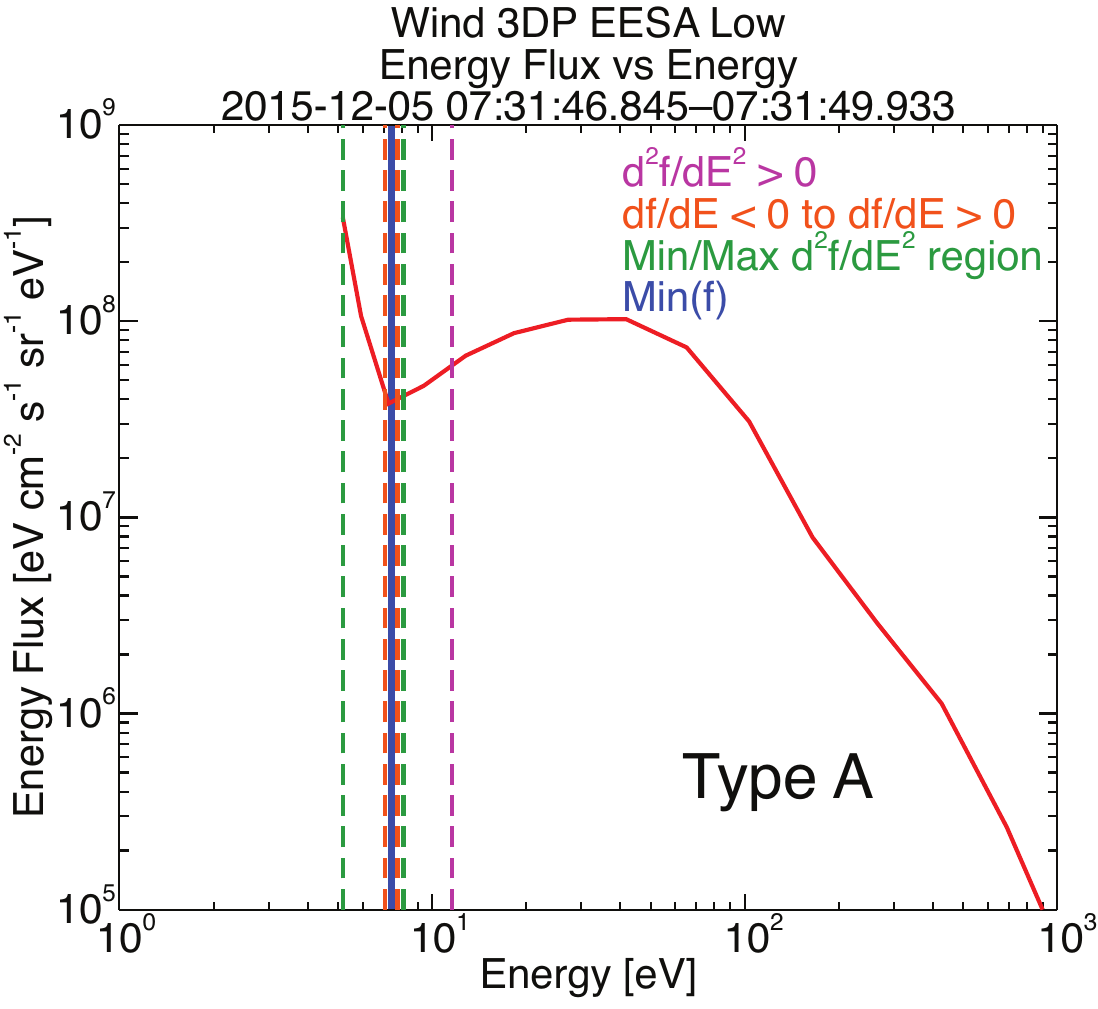}}
    \caption{Illustrative example of an electron energy distribution function (EDF) showing the photoelectrons and the ambient electrons in the solar wind near 1 AU.  The red solid line shows the median of the energy flux, $f{\scriptstyle_{e}}\left( E \right)$ [$eV^{+1} \ cm^{-2} \ s^{-1} \ sr^{-1} \ eV^{-1}$], at each energy over all 88 solid angle bins versus energy, $E$ [$eV$], on a log-log scale.  The vertical lines indicate values or ranges of values for the estimated $\phi{\scriptstyle_{sc}}$ of this EDF (see labels).  The magenta lines (\textbf{Method 1}, see text for definition) show the region bounding the region of positive curvature (note left-hand bound covered by green dashed line).  The orange lines show the region bounding the transition from negative to positive slope (\textbf{Method 2}).  The green lines show the region bounding the minimum-to-maximum curvature transition (\textbf{Method 3}).  Finally, the blue line shows the point of minimum in $f{\scriptstyle_{e}}\left( E \right)$ (\textbf{Method 4}).}
    \label{fig:ExampleEDF}
\end{figure}
%%++++++++++++++++++++++++++++++++++++++++++++++++++++++++++++++++++++++++++++++++++++++++
%% Image:  Example EDF
%%++++++++++++++++++++++++++++++++++++++++++++++++++++++++++++++++++++++++++++++++++++++++

\indent  Figure \ref{fig:ExampleEDF} shows an example EDF to illustrate how we identify the region in energy filled by photoelectrons versus those with ambient electrons.  The data in Figure \ref{fig:ExampleEDF} (solid red line) show the median at each energy of all the thin, multi-colored lines (i.e., the 88 solid angle bins) from Figure \ref{fig:ExamplePhotoelectrons} but in units of energy flux\footnote{Energy flux was used when calculating $\phi{\scriptstyle_{sc}}$ as the changes in $f{\scriptstyle_{e}}\left( E \right)$ tend to be more pronounced than in units of phase space density.}.  For brevity, we will call this $f{\scriptstyle_{med}}\left( E \right)$.  There are also multiple vertical lines in Figure \ref{fig:ExampleEDF} that will be explained below.  Figure \ref{fig:ExampleEDF} is an ideal case where there is a clear discontinuity in the data, below which one sees a sharply increasing $f{\scriptstyle_{e}}\left( E \right)$ with decreasing $E$.  As explained in the discussion above and referenced citations, this ``kink'' in the EDF roughly marks the point corresponding to $\phi{\scriptstyle_{sc}}$.

%%****************************************************************************************
%%  Subsection:  Floating Potential Calculation
%%****************************************************************************************
\phantomsection   %%  Fix reference link
\subsection{Floating Potential Calculation}  \label{subsec:FloatingPotentialCalculation}

\indent  The EESA Low detector returns measurements for 88 solid angle look directions and 15 energy bins.  To simplify and reduce computation requirements, we construct a pitch-angle distribution (PAD) for each EDF where we average the solid angle look directions within $\pm$22.5$^{\circ}$ of the parallel, perpendicular, and anti-parallel directions (with respect to the quasi-static magnetic field vector, $\mathbf{B}{\scriptstyle_{o}}$).  This reduces the number of computations for each EDF by a factor of $\sim$30.  Note that when we examine an arbitrary EDF and view all 88 solid angles, the profiles/shapes are nearly always the same.  That is, if one solid angle direction looks like the EDF in Figure \ref{fig:ExampleEDF}, the other 87 will as well.

\indent  To keep the approach simple and focus on finding accurate but quick solutions for $\phi{\scriptstyle_{sc}}$, we use four methods to determine $\phi{\scriptstyle_{sc}}$, which is a rough boundary between photoelectron and ambient electrons.  Note that each of the four $\phi{\scriptstyle_{sc}}$ methods illustrated in Figure \ref{fig:ExampleEDF} will have at least three values.  The four methods are as follows:

\begin{description}[itemsep=0pt,parsep=0pt,topsep=0pt]
  \item[Method 1]  find the range of energies where $\tfrac{ d^{2} f }{ dE^{2} }$ $>$ 0, or the positive curvature region (dashed magenta lines in Figure \ref{fig:ExampleEDF});
  \item[Method 2]  find the range of energies where $\tfrac{ d f }{ dE }$ transitions from negative to positive, i.e., find the local minimum of $f{\scriptstyle_{e}}\left( E \right)$ (dashed orange lines in Figure \ref{fig:ExampleEDF});
  \item[Method 3]  find the range of energies bounding the minimum and maximum values of $\tfrac{ d^{2} f }{ dE^{2} }$, i.e., region of minimum and maximum curvature (dashed green lines in Figure \ref{fig:ExampleEDF}); and
  \item[Method 4]  find the local minimum in $f{\scriptstyle_{e}}\left( E \right)$ between $E{\scriptstyle_{min}}$ and $E{\scriptstyle_{max}}$ (solid blue line in Figure \ref{fig:ExampleEDF}).
\end{description}

%%++++++++++++++++++++++++++++++++++++++++++++++++++++++++++++++++++++++++++++++++++++++++
%% Image:  Example EDF Shapes
%%++++++++++++++++++++++++++++++++++++++++++++++++++++++++++++++++++++++++++++++++++++++++
\begin{figure}
  \centering
    {\includegraphics[trim = 0mm 0mm 0mm 0mm, clip, width=80mm]{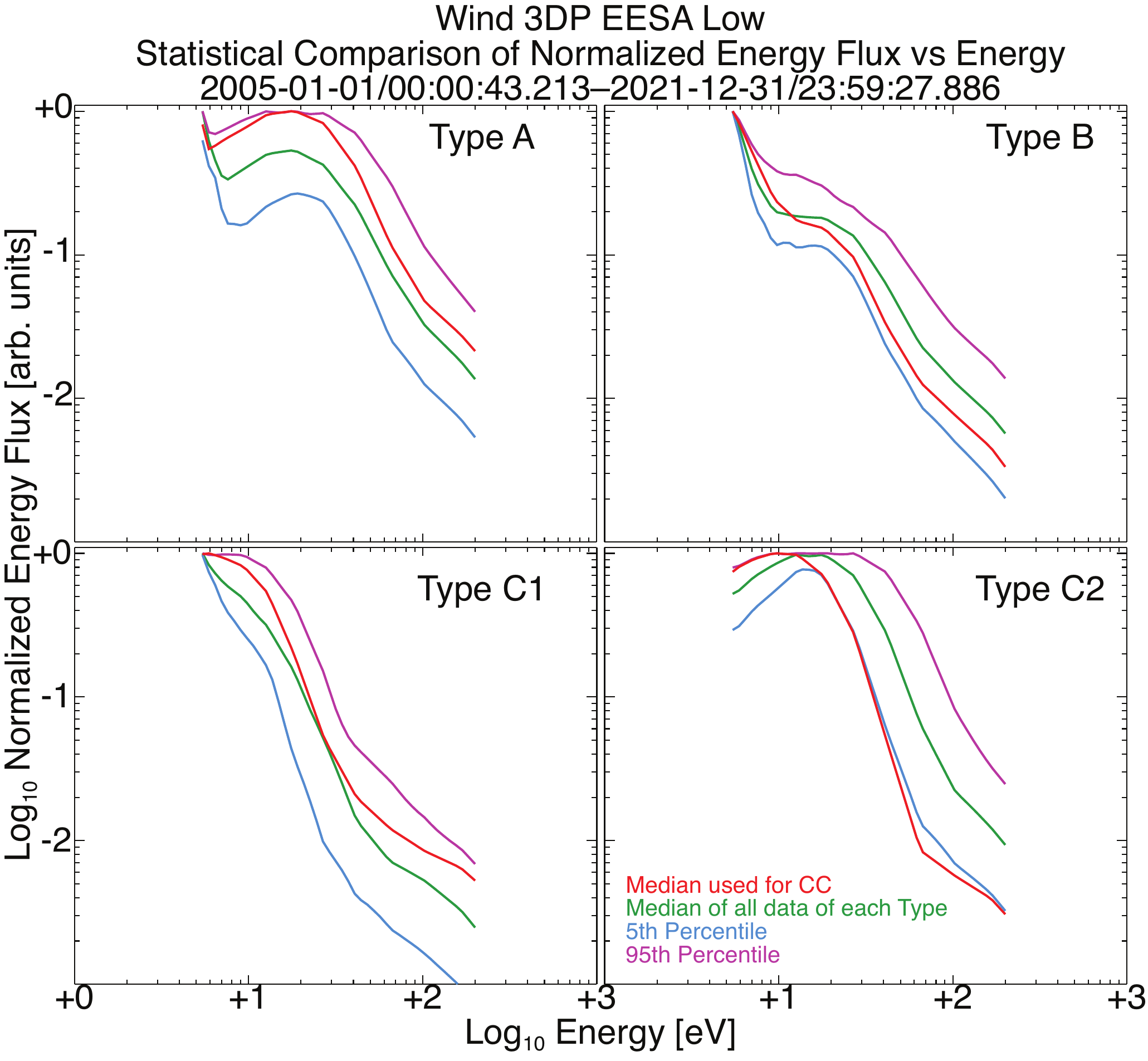}}
    \caption{Characteristic EDFs representing the four shapes observed herein, shown as the red solid lines.  All $f{\scriptstyle_{e}}\left( E \right)$ were normalized to the peak magnitude prior to calculation for this figure.  Type A is the ideal case, illustrated in Figure \ref{fig:ExampleEDF}.  Type B is the next best shape but often shows a shoulder or saddle-point rather than a local minimum, increasing the uncertainty of the $\phi{\scriptstyle_{sc}}$ estimates.  Type C1 has $\phi{\scriptstyle_{sc}}$ $<$ $E{\scriptstyle_{min}}$ and the lowest energy bin has $\tfrac{ d f }{ dE }$ $\sim$ 0.  Type C2 also has $\phi{\scriptstyle_{sc}}$ $<$ $E{\scriptstyle_{min}}$ but the lowest energy bin has $\tfrac{ d f }{ dE }$ $>$ 0.  Both Type C1 and C2 are not considered useful shapes as we can only constrain the upper bound on $\phi{\scriptstyle_{sc}}$.  The green line in each panel represents the median of all EDFs for each Type.  The blue(magenta) line in each panel represents the corresponding 5$^{th}$(95$^{th}$) percentile of all EDFs of each type.}
    \label{fig:FourShapes}
\end{figure}
%%++++++++++++++++++++++++++++++++++++++++++++++++++++++++++++++++++++++++++++++++++++++++
%% Image:  Example VDF Shapes
%%++++++++++++++++++++++++++++++++++++++++++++++++++++++++++++++++++++++++++++++++++++++++

%%****************************************************************************************
%%  Subsection:  Assigning Type Labels
%%****************************************************************************************
\phantomsection   %%  Fix reference link
\subsection{Assigning Type Labels}  \label{subsec:AssigningTypeLabels}

\indent  Figure \ref{fig:FourShapes} shows the EDF profile of the four characteristic shapes we defined in this study (all lines have been normalized by their respective peak values).  During the examination of the EDFs, we observed two useful shapes that satisfy $E{\scriptstyle_{min}}$ $<$ $\phi{\scriptstyle_{sc}}$ (i.e., Types A and B in Figure \ref{fig:FourShapes}) and two not so useful shapes/profiles that satisfy $E{\scriptstyle_{min}}$ $>$ $\phi{\scriptstyle_{sc}}$ (i.e., Types C1 and C2 in Figure \ref{fig:FourShapes}).  We chose, at random, one example EDF for each of the four characteristic shapes.  We then calculate $f{\scriptstyle_{med}}\left( E \right)$ (i.e., same as done for Figure \ref{fig:ExampleEDF}) to construct an example type, shown as the red lines in each panel of Figure \ref{fig:FourShapes}.  We then compared these example types to all \totalnvdfsall~EDFs to assign the type labels accordingly (method for assignment discussed below).  After the EDFs were assigned a type label, we needed to further validate that the assignments are correct.  We calculated one-variable statistics of $f{\scriptstyle_{med}}\left( E \right)$ at each energy for all EDFs in each type category\footnote{One can think of this similar to a superposed epoch analysis but performed on EDFs.  It allows us to statistically examine the EDFs from the resulting labeling scheme without checking each EDF ``by eye'' to ensure they exhibit the proper shape.}.  For instance, for all EDFs labeled Type A, there will be N values of $f{\scriptstyle_{med}}\left( E{\scriptstyle_{j}} \right)$ at energy $E{\scriptstyle_{j}}$.  We can calculate the one-variable statistics of $f{\scriptstyle_{med}}\left( E{\scriptstyle_{j}} \right)$ to get things like the mean ($\bar{X}$), median ($\tilde{X}$), 5$^{th}$ percentile ($X{\scriptstyle_{5\%}}$), 95$^{th}$ percentile ($X{\scriptstyle_{95\%}}$), etc. (see Appendix \ref{app:Definitions} for symbol definitions).  Then for, say, $X{\scriptstyle_{5\%}}$ we can construct a line of 15 points (one for each energy) for this specific one-variable statistic value.  We can do this for any of the one-variable statistic values.  We chose the values of $X{\scriptstyle_{5\%}}$, $\tilde{X}$, and $X{\scriptstyle_{95\%}}$, which are shown as the blue, green, and magenta lines, respectively, in Figure \ref{fig:FourShapes}.  As one can see, the profile of the blue, green, and magenta lines for each EDF type follows that of the characteristic example used to assign the type label.  Thus, the assignments are validated as being statistically correct.

\indent  We needed a quick method for identifying all the EDFs with an associated type label.  This is necessary to determine which of the \totalnvdfsall~values of $\phi{\scriptstyle_{sc}}$ we can trust.  The quickest approach is to calculate the cross-correlation coefficient (CC) between each characteristic example $f{\scriptstyle_{med}}\left( E \right)$ (shown as red lines in Figure \ref{fig:FourShapes}) and the $f{\scriptstyle_{med}}\left( E \right)$ for all \totalnvdfsall~EDFs.  We do this in the following steps:
\begin{enumerate}[itemsep=0pt,parsep=0pt,topsep=0pt]
  \item  normalize all \totalnvdfsall~$f{\scriptstyle_{med}}\left( E \right)$ by their respective peak values to avoid vertical offsets when comparing to each characteristic example $f{\scriptstyle_{med}}\left( E \right)$ (also normalized by peak value);
  \item  calculate \totalnvdfsall~CCs for each characteristic example $f{\scriptstyle_{med}}\left( E \right)$ allowing for seven energy offsets (i.e., allows each EDF to shift left and right in energy), generating a [N,7]-element array of CCs for each characteristic example $f{\scriptstyle_{med}}\left( E \right)$;
  \item  find the maximum CC by energy offset then by characteristic example $f{\scriptstyle_{med}}\left( E \right)$ to determine the associated shape/type for each EDF; and
  \item  then ensure that the slope\footnote{Note that this slope requirement supersedes the CC magnitude ranking.} of the first few energy bins for Types A, B, and C2 match that expected from the characteristic example $f{\scriptstyle_{med}}\left( E \right)$ (i.e., Types A and B must have a negative slope, Type C2 must have a positive slope).
\end{enumerate}

\noindent  We did not bother with further constraints or complications because as we will discuss below, there are many fewer Type C1 and Type C2 EDFs than Type A or B.  The one-variable statistics of all EDFs for each type shown as the blue, green, and magenta lines in Figure \ref{fig:FourShapes} indicate that our type label assignments are statistically correct.

\indent  The one-variable statistics of all CC values for each type are shown below in the following form $X{\scriptstyle_{5\%}}$--$X{\scriptstyle_{95\%}}$($\bar{X}$)[$\tilde{X}$] (see Appendix \ref{app:Definitions} for symbol definitions)

\begin{description}[itemsep=0pt,parsep=0pt,topsep=0pt]
  \item[Type A]         0.9915--0.9998(0.9965)[0.9969];
  \item[Type B]         0.9918--0.9998(0.9968)[0.9975];
  \item[Type C1]        0.9864--0.9996(0.9957)[0.9974];
  \item[Type C2]        0.9832--0.9997(0.9933)[0.9944];
  \item[Type C1 \& C2]  0.9833--0.9997(0.9938)[0.9955].
\end{description}

\begin{figure}
  \centering
    {\includegraphics[trim = 0mm 0mm 0mm 0mm, clip, width=80mm]{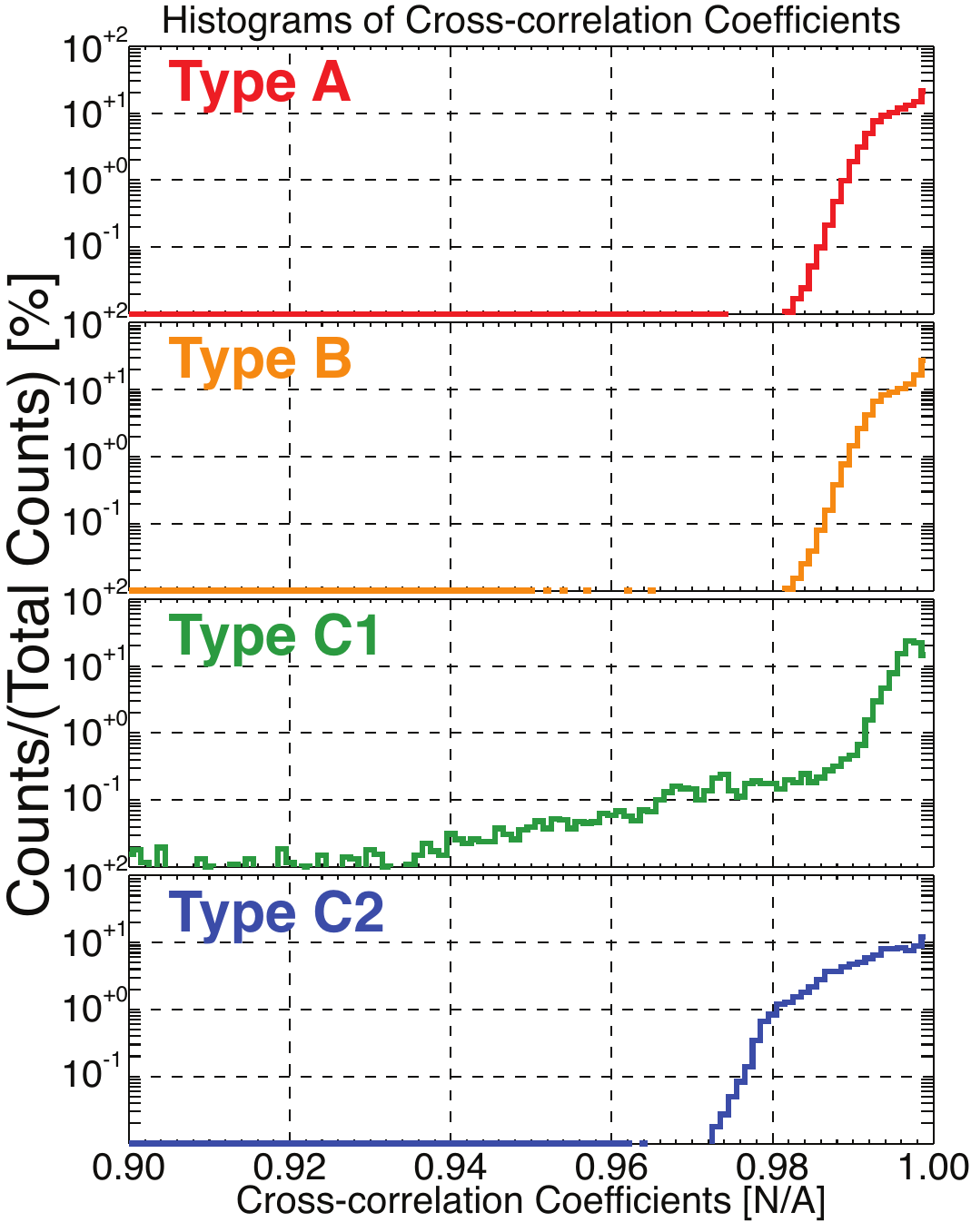}}
    \caption{Histograms of the CC values for each type are shown as percentages of the total number of counts for each type.}
    \label{fig:CCHistograms}
\end{figure}

\indent  Figure \ref{fig:CCHistograms} shows the one-dimensional histograms of the CCs for each type, illustrating they have distinct distributions.  For both Types A and B, the distribution is a narrow peak with $\sim$95\% of all CCs above $\sim$0.991 and very weak (if any) tails to lower CC magnitudes.  Both Type C1 and C2 had much broader peaks and stronger tails extending to lower CC magnitudes (especially C1).  This further validates the identification scheme discussed above.

%%++++++++++++++++++++++++++++++++++++++++++++++++++++++++++++++++++++++++++++++++++++++++
%% Image:  1D Histograms of CCs
%%++++++++++++++++++++++++++++++++++++++++++++++++++++++++++++++++++++++++++++++++++++++++
%%++++++++++++++++++++++++++++++++++++++++++++++++++++++++++++++++++++++++++++++++++++++++
%% Image:  1D Histograms of CCs
%%++++++++++++++++++++++++++++++++++++++++++++++++++++++++++++++++++++++++++++++++++++++++

\indent  Thus, our labeling scheme identified, from all \totalnvdfsall \ EDFs, that there are:

\begin{description}[itemsep=0pt,parsep=0pt,topsep=0pt]
  \item[Type A]         \totalnvdfsAs(\totalpvdfsAs \%) EDFs;
  \item[Type B]         \totalnvdfsBs(\totalpvdfsBs \%) EDFs;
  \item[Type C1]        \totalnvdfsCo(\totalpvdfsCo \%) EDFs;
  \item[Type C2]        \totalnvdfsCt(\totalpvdfsCt \%) EDFs; and
  \item[Type C1 \& C2]  \totalnvdfsCs(\totalpvdfsCs \%) EDFs.
\end{description}

\noindent  Given the small number of Type C1 and C2 EDFs, we did not further enhance the differentiation algorithms.  The statistical shapes shown in Figure \ref{fig:FourShapes} illustrate that our algorithm works and that we have properly classified the EDFs.  Therefore, we found no need to refine further.

\indent  We define a discrete quality flag for each set of $\phi{\scriptstyle_{sc}}$ as follows:

\begin{description}[itemsep=0pt,parsep=0pt,topsep=0pt]
  \item[QF $=$ 4]  All EDFs of Type A;
  \item[QF $=$ 2]  All EDFs of Type B; and
  \item[QF $=$ 0]  All EDFs of Type C1 or C2.
\end{description}

\noindent  Again, this is is to simplify the process and use of the dataset.  It is to help the user determine whether to use or not use any given $\phi{\scriptstyle_{sc}}$ solution.  We did not feel it was useful to delineate the quality flags further as all Type Cs are one-sided constrained and all Type As are good.  Type B solutions are reliable and most are good but a few had weaker saddle points than others, thus we used a less confident QF than for Type A.

%%----------------------------------------------------------------------------------------
%%  Section:  Solar Wind Statistics
%%----------------------------------------------------------------------------------------
\phantomsection   %%  Fix reference link
\section{Solar Wind Statistics}  \label{sec:SolarWindStatistics}

\indent  In this section we present some preliminary statistical analysis of our $\phi{\scriptstyle_{sc}}$ estimates.

\indent  Table \ref{tab:AllEDFs} shows the one-variable statistics of $\phi{\scriptstyle_{sc}}$ for all \totalnvdfsall~EDFs, i.e., not distinguishing by EDF type\footnote{We also list the one-variable statistics for the methods separated by EDF type in Appendix \ref{app:ExtraStatistics}.}.  Note that the $\tilde{X}$ values for \textbf{Method 2} and \textbf{Method 3} nicely bound the values for \textbf{Method 4}.  \textbf{Method 1} has a much broader range of values but this is not surprising as \textbf{Method 1} can be ill-constrained except for EDF Types A and C2.  The main takeaway here is that the spacecraft potential for the \emph{Wind} spacecraft near 1 AU in the solar wind typically satisfies 5 eV $\lesssim$ $\phi{\scriptstyle_{sc}}$ $\lesssim$ 13 eV, consistent with previous studies \citep[e.g.,][]{geach05a, pedersen95a, salem01a}.

\begin{figure}
  \centering
    {\includegraphics[trim = 0mm 0mm 0mm 0mm, clip, width=80mm]{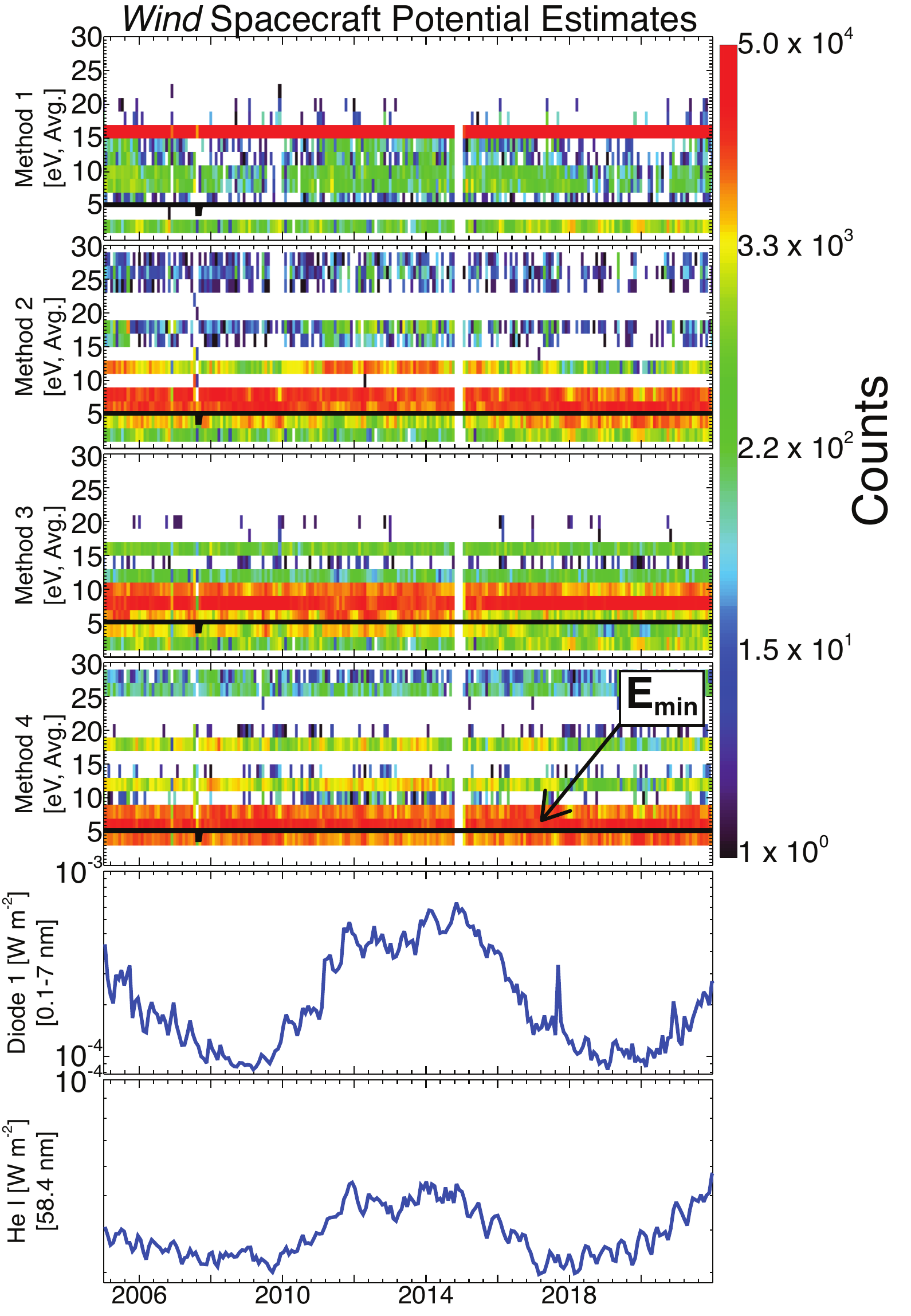}}
    \caption{2D histograms of the $\phi{\scriptstyle_{sc}}$ estimates for \textbf{Method 1}--\textbf{Method 4} (shown in order).  The thick black line in each panel shows the $E{\scriptstyle_{min}}$ value versus time.  The horizontal (time) axis is parsed into four week bins while the vertical axis is parsed into 16 uniformly sized bins between 0 eV and 30 eV (i.e., roughly 2 eV each).  The color scale shows total counts in each 2D histogram bin.  The bottom two panels show solar irradiances [$W \ m^{-2}$] measured with two different detectors from the TIMED SEE experiment.  These are shown as proxies for solar cycle variation and ionizing photon intensities versus time.}
    \label{fig:2DHist}
\end{figure}

\indent  Figure \ref{fig:2DHist} shows the 2D histograms of the $\phi{\scriptstyle_{sc}}$ estimates for \textbf{Method 1}--\textbf{Method 4} (shown in order).  The histogram bin sizes are four weeks for the horizontal (time) axis and $\sim$2 eV (i.e., 16 uniformly sized bins between 0 eV and 30 eV) for the vertical axis.  White bins indicate no data whereas colors are quantified in the logarithmic scale to the right.  Note that blue and purple are only a few tens of counts to single digits whereas the yellow-to-red portion are in the thousands.  Thus, the latter are statistically significant while the former have low signal-to-noise ratios.

\indent  Recall that \textbf{Method 1}--\textbf{Method 3} result in a range of values, i.e., two per EDF.  Therefore, to construct these histograms for \textbf{Method 1}--\textbf{Method 3}, we calculate the median over the three PAD directions first, then take the average of the range of values.  For \textbf{Method 4}, we take the median from the three pitch-angles and then construct the histogram.  The bottom two panels show solar irradiance measurements over the wavelength range of 0.1--7.0 nm (Diode 1) and $\sim$58.4 nm (He I) observed by the TIMED SEE experiment.

%%========================================================================================
%%  Table:  All EDFs
%%========================================================================================
\startlongtable  %%  In case table runs long
\begin{deluxetable}{| l | c | c | c | c | c | c |}
  \tabletypesize{\footnotesize}    %%  Table font size
  %%  Table Caption
  \tablecaption{All EDF Types \label{tab:AllEDFs}}
  %%  Column Headers
  \tablehead{\colhead{$\phi{\scriptstyle_{sc}}$ [eV]} & \colhead{$X{\scriptstyle_{min}}$}\tablenotemark{a} & \colhead{$X{\scriptstyle_{5\%}}$}\tablenotemark{b} & \colhead{$\bar{X}$}\tablenotemark{c} & \colhead{$\tilde{X}$}\tablenotemark{d} & \colhead{$X{\scriptstyle_{95\%}}$}\tablenotemark{e} & \colhead{$X{\scriptstyle_{max}}$}\tablenotemark{f}}
  %%======================================================================================
  %%  Table Body
  %%======================================================================================
  \startdata
  \multicolumn{7}{ |c| }{\textit{All \totalnvdfsall~EDFs, Lower Bound}} \\
  \hline
  %%                    Min    5%    Avg    Med    95%     Max
  \textbf{Method 1}  & 1.00 & 5.18 & 5.05 & 5.18 & 5.18 & 17.3  \\
  \textbf{Method 2}  & 1.00 & 5.42 & 7.98 & 7.09 & 12.1 & 27.1  \\
  \textbf{Method 3}  & 1.00 & 5.18 & 5.24 & 5.18 & 5.93 & 16.6  \\
  \textbf{Method 4}  & 4.91 & 5.93 & 8.03 & 7.41 & 12.7 & 29.7  \\
  \hline
  \multicolumn{7}{ |c| }{\textit{All \totalnvdfsall~EDFs, Upper Bound}} \\
  \hline
  %%                    Min    5%    Avg    Med    95%     Max
  \textbf{Method 1}  & 4.91 & 16.6 & 28.5 & 29.7 & 29.7 & 29.7  \\
  \textbf{Method 2}  & 4.91 & 5.93 & 8.88 & 7.75 & 13.3 & 29.7  \\
  \textbf{Method 3}  & 4.91 & 5.93 & 12.7 & 12.7 & 18.1 & 29.7  \\
  \textbf{Method 4}  & 4.91 & 5.93 & 8.92 & 7.41 & 18.1 & 29.7  \\
  \hline
  \enddata
  %%======================================================================================
  %%  Table Notes
  %%======================================================================================
  \tablenotetext{a}{minimum}
  \tablenotetext{b}{5$^{th}$ percentile}
  \tablenotetext{c}{mean}
  \tablenotetext{d}{median}
  \tablenotetext{e}{95$^{th}$ percentile}
  \tablenotetext{f}{maximum}
  \tablecomments{For symbol definitions, see Appendix \ref{app:Definitions}.}
\vspace{-20pt}
\end{deluxetable}
%%========================================================================================
%%  Table:  All EDFs
%%========================================================================================

\indent  Note that \textbf{Method 1} has, by construction, the largest range of values and often is bounded by our the instrument $E{\scriptstyle_{min}}$ and our assumed $E{\scriptstyle_{max}}$ values.  For EDFs of Type C1 and C2, the lower bound for \textbf{Method 1}--\textbf{Method 3} is often 1 eV\footnote{This was a default minimum we chose based upon the physically justified (and empirically motivated) assumption that the spacecraft will not charge negative in the solar wind in sunlight.}.  The midpoint over the range of energies allowed is 15 eV.  This is why there is a high density of counts near 15 eV for \textbf{Method 1}.  Since these are constructed using four week bins, each column can have multiple values, but this does not imply structure in the photoelectron part of any individual EDF.  That is, finite counts in multiple rows of the histograms indicate that during any given four week interval (i.e., one column), $\phi{\scriptstyle_{sc}}$ varied a lot.

\indent  For EDFs of Type A, \textbf{Method 2}--\textbf{Method 4} are consistent with \textbf{Method 2} and \textbf{Method 3} bounding \textbf{Method 4}.  This is largely why these three methods show a strong peak below $\sim$10 eV for all times.  For EDFs of Type B, \textbf{Method 2} and \textbf{Method 4} are in the best agreement while \textbf{Method 3} has difficulty limiting the bounds of $\phi{\scriptstyle_{sc}}$.  Note that the repeated value of $\sim$4.91 eV in Table \ref{tab:AllEDFs} corresponds to $\sim$95\% of most $E{\scriptstyle_{min}}$ detector values.  This was a default guess solution in the software when it was clear that $E{\scriptstyle_{min}}$ $>$ $\phi{\scriptstyle_{sc}}$ (i.e., for EDFs of Type C1 and C2).  For reference, previous work has shown that typical $\phi{\scriptstyle_{sc}}$ magnitudes for \emph{Wind} in the solar wind are below $\sim$10 eV \citep[e.g.,][]{salem01a, wilsoniii19a}, consistent with the $\phi{\scriptstyle_{sc}}$ estimates for \textbf{Method 2}--\textbf{Method 4} shown in Figure \ref{fig:2DHist}.

\indent  The $\phi{\scriptstyle_{sc}}$ estimates in Figure \ref{fig:2DHist} also show an interesting solar cycle variation, consistent with other studies (e.g., \textit{Salem et al.,} in preparation).  We focus on the peaks in the yellow-to-red color range as they have statistically significant counts.  The histograms for \textbf{Method 2} (second panel from top) show a bifurcated structure\footnote{That is, there are two peaks in the same time bin.} in $\phi{\scriptstyle_{sc}}$ that varies in time (\textbf{Method 4} shows this as well, but to a weaker extent).  The first bifurcated peak occurs before 2006, which corresponds to the tail end of solar cycle 23 (e.g., see trend in Diode 1 panel).  The second bifurcated peak starts just before 2012 and extends to nearly 2016, which correponds to the peak in sunspot numbers (SSNs) for solar cycle 24 (also illustrated in trend in Diode 1 panel).  Note that solar cycle 24 had a double-peaked structure in SSNs, with the first peak just before 2012 and the second in early 2014.  The last three years of the data shown in Figure \ref{fig:2DHist} are the start of solar cycle 25 corresponding to solar minimum, thus only a weak bifurcated peak is starting to show.  Note that there is some hint of a third peak for \textbf{Method 2} and \textbf{Method 4} in Figure \ref{fig:2DHist}.

\indent  There are two environments known to enhance $\phi{\scriptstyle_{sc}}$ in the solar wind, low $n{\scriptstyle_{e}}$ regions and the sheath regions downstream of collisionless shock waves \citep[e.g.,][]{wilsoniii19a}.  The low $n{\scriptstyle_{e}}$ regions have lower ambient electron currents which increases the relative importance of the escaping photoelectron current, thus increasing $\phi{\scriptstyle_{sc}}$ to more positive values.  The sheath regions downstream of shock waves have much stronger ion currents, which increase $\phi{\scriptstyle_{sc}}$ to more positive values.  Interplanetary (IP) shock waves driven by coronal mass ejections (CMEs) \citep[e.g.,][]{jian06b, richardson18a} and/or stream interaction regions (SIRs) \citep[e.g.,][]{jian06a, richardson18a} are more common during solar maximum than solar minimum, thus show a clear solar cycle variation.  Thus, the bifurcated peak likely occurs because of IP shock waves.  As stated above, we are not implying that the solar irradiance changes versus time directly result in larger $\phi{\scriptstyle_{sc}}$.  Rather, the solar irradiance measures are proxies for solar cycle variations which directly influence the rate of CME and SIR occurrence.

\indent  The magnitude of $\phi{\scriptstyle_{sc}}$ is observed to increase downstream of IP shocks \citep[e.g.,][]{wilsoniii19a}, which is likely due to increases in the ion currents\footnote{That is, the ion temperature typically increases much more than the electron temperature in shock sheath regions, which increases the ion thermal currents more than the electron.} relative to the electron thermal and photoelectron currents.  This drives the spacecraft more electrically positive as it reaches a new current balanced equilibrium.  It may also be related to enhanced secondary electron generation internal to the instrument as seen on other spacecraft in shock sheaths \citep[e.g.,][]{gershman17a}, but this is beyond the scope of the present study.

%%++++++++++++++++++++++++++++++++++++++++++++++++++++++++++++++++++++++++++++++++++++++++
%% Image:  2D Histograms
%%++++++++++++++++++++++++++++++++++++++++++++++++++++++++++++++++++++++++++++++++++++++++
%%++++++++++++++++++++++++++++++++++++++++++++++++++++++++++++++++++++++++++++++++++++++++
%% Image:  2D Histograms
%%++++++++++++++++++++++++++++++++++++++++++++++++++++++++++++++++++++++++++++++++++++++++

\indent  As discussed before, \textbf{Method 1}--\textbf{Method 3} provide a range of values while \textbf{Method 4} is a single value.  However, all of these have three solutions for each of the three pitch-angle ranges discussed in Section \ref{sec:DefinitionsDataSets}.  To provide a range of values for \textbf{Method 4}, we take the minimum and maximum values of $E$ from the three pitch-angles.  We can then calculate one-variable statistics for each of the methods for all EDFs or by type and provide a range of values for each one-variable statistic variable (e.g., this is in reference to the values shown in Table \ref{tab:AllEDFs}).

%%++++++++++++++++++++++++++++++++++++++++++++++++++++++++++++++++++++++++++++++++++++++++
%% Image:  FFT of all phi_sc
%%++++++++++++++++++++++++++++++++++++++++++++++++++++++++++++++++++++++++++++++++++++++++
\begin{figure}
  \centering
    {\includegraphics[trim = 0mm 0mm 0mm 0mm, clip, width=80mm]{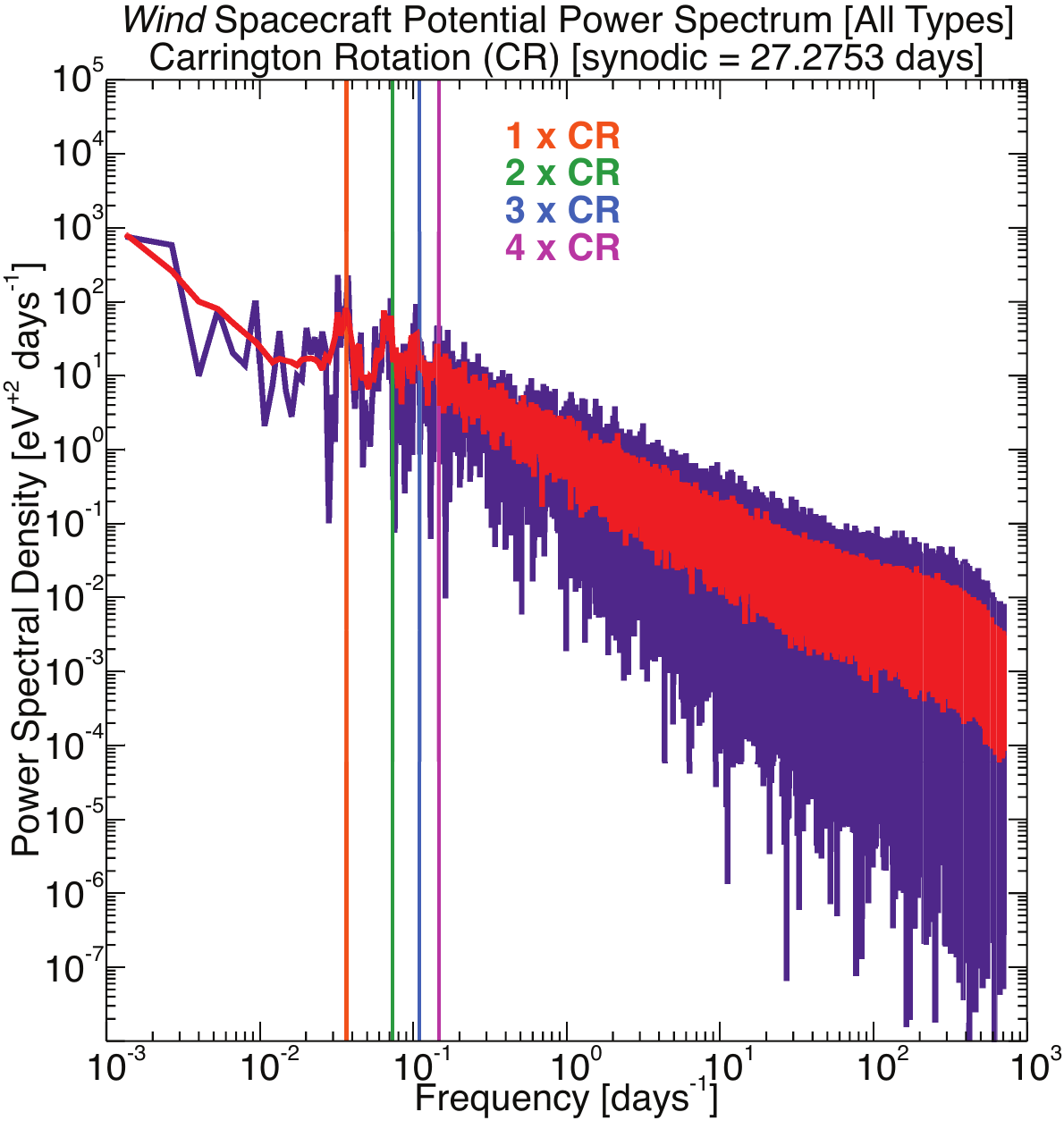}}
    \caption{The power spectral density (PSD) [$eV^{+2} \ days^{-1}$] of the fast Fourier transform (FFT) of all $\phi{\scriptstyle_{sc}}$ estimates for all four EDF types, using \textbf{Method 4} (note this is the median of the three PADs), versus frequency [$days^{-1}$] shown in purple.  The red line is the smoothed PSD, using a running boxcar average with an eight point bin width.  Overplotted are four vertical lines corresponding to the frequency of the synodic Carrington rotation period ($\sim$27.2753 days) and its first three harmonics.}
    \label{fig:FFTofphisc}
\end{figure}
%%++++++++++++++++++++++++++++++++++++++++++++++++++++++++++++++++++++++++++++++++++++++++
%% Image:  FFT of all phi_sc
%%++++++++++++++++++++++++++++++++++++++++++++++++++++++++++++++++++++++++++++++++++++++++

\indent  The main takeaway from Table \ref{tab:AllEDFs} and Figure \ref{fig:2DHist} is that for the majority of the mission, the following is satisfied 5 eV $\lesssim$ $\phi{\scriptstyle_{sc}}$ $\lesssim$ 13 eV, with peak occurrence rates in the $\sim$6--9 eV range.  In fact, most of the intervals where $\phi{\scriptstyle_{sc}}$ $\gtrsim$ 9 eV correspond to the sheath regions of interplanetary shocks \citep[e.g., see discussion in][]{wilsoniii19a} because the ion thermal currents are much larger than in the ambient solar wind.

\indent  Figure \ref{fig:FFTofphisc} shows the PSD of all $\phi{\scriptstyle_{sc}}$ estimates for all four EDF types (using \textbf{Method 4}) versus frequency.  We also superposed the corresponding frequency of the synodic Carrington rotation period ($\sim$27.2753 days) and its first three harmonics.  One can immediately see that $\phi{\scriptstyle_{sc}}$ peaks at this frequency and its harmonics.  The most likely explanation is due to the variations in $j{\scriptstyle_{th,s}}$ and $j{\scriptstyle_{o,s}}$ caused by SIRs \citep[e.g.,][]{jian06a, richardson18a}, which typically occur at intervals defined roughly by the Carrington rotation of the Sun.  We examined both the synodic ($\sim$27.2753 days) and sidereal ($\sim$25.38 days), periods but the one of relevance is the synodic.  An interesting observation is that these peaks are much less well defined when we construct the PSD using only Types A and B and are completely missing if we look at Type C1 or C2 or both C1 and C2.  It is not immediately clear why the peaks change in amplitude or disappear altogether when isolating the different EDF types.  However, this is beyond the scope of the current study.

%%----------------------------------------------------------------------------------------
%%  Section:  Discussion and Conclusions
%%----------------------------------------------------------------------------------------
\phantomsection   %%  Fix reference link
\section{Discussion and Conclusions}  \label{sec:DiscussionandConclusions}

\indent  We present measurements of the spacecraft floating potential for the \emph{Wind} spacecraft near 1 AU in the solar wind between January 1, 2005 and January 1, 2022 (i.e., 17 years).  The analysis was performed as the first step in an effort to properly calibrate the \emph{Wind} 3DP thermal electron detector, EESA Low.  We describe the resulting publicly available dataset and some preliminary results of the long-term statistical trends.

\indent  The long-term, statistical estimate for $\phi{\scriptstyle_{sc}}$ satisfies 5 eV $\lesssim$ $\phi{\scriptstyle_{sc}}$ $\lesssim$ 13 eV, with peaks in the $\sim$6--9 eV range, consistent with previous measurements in the solar wind \citep[e.g.,][]{salem01a}.  Note that part of the reason for the small magnitudes of $\phi{\scriptstyle_{sc}}$ is that \emph{Wind} was designed and built with strict requirements for electromagnetic and electrostatic cleanliness \citep[e.g., see][and references therein]{wilsoniii21b}.  This is an important consideration in mission development and instrument calibration, as less stringent requirements lead to larger $\phi{\scriptstyle_{sc}}$, which will result in larger deviations between the observed energy-angle of an incident particle versus its true energy-angle \citep[e.g.,][]{barrie19a}.

\indent  When we examined the statistical trends of $\phi{\scriptstyle_{sc}}$ versus time, we found a bifurcated peak that correlated with solar maximum\footnote{A 1D cut through the second peak for \textbf{Method 1} in Figure \ref{fig:2DHist} shows the count rates of this higher peak vary in phase with the solar irradiance measures shown at the bottom of Figure \ref{fig:2DHist}, i.e., they are highly correlated.}.  This is most likely due to enhanced ion currents downstream of IP shocks and/or regions of high ion temperature and low total electron density.  We also found peaks in the power spectral density at the frequency corresponding to the synodic Carrington rotation period ($\sim$27.2753 days) and its first three harmonics.  The obvious candidates that depend upon Carrington rotation and solar cycle are SIRs and CMEs.

\indent  A critical takeaway from this study is that future thermal electron instrument design should require that $E{\scriptstyle_{min}}$ $<$ $\phi{\scriptstyle_{sc}}$ (for positive $\phi{\scriptstyle_{sc}}$) for the largest fraction of the measurements possible in any given region of space.  This is critical for two reasons.  The first is that when $E{\scriptstyle_{min}}$ $<$ $\phi{\scriptstyle_{sc}}$, the instrument is measuring all ambient electrons within its field-of-view as they are accelerated to energies of at least $e \phi{\scriptstyle_{sc}}$ before entering the detector (for positive $\phi{\scriptstyle_{sc}}$).  If $E{\scriptstyle_{min}}$ $>$ $\phi{\scriptstyle_{sc}}$, the instrument does not measure all ambient electrons and, in fact, misses most of the velocity distribution.  This can result in highly inaccurate velocity moments and affect the interpretation of any analysis performed on the data \citep[e.g.,][]{song97a, wilsoniii22h}.  There are ways to estimate $\phi{\scriptstyle_{sc}}$ when $E{\scriptstyle_{min}}$ $>$ $\phi{\scriptstyle_{sc}}$ as previously discussed \citep[e.g., see][]{salem01a, salem23a}.  However, it requires multiple assumptions and very well calibrated instrumentation.

\indent  The second reason one should require that $E{\scriptstyle_{min}}$ $<$ $\phi{\scriptstyle_{sc}}$ is that it gives at least one direct, independent measure of $\phi{\scriptstyle_{sc}}$.  That is, the observed ``kink-like'' change in the EDF as illustrated in the example in Figure \ref{fig:ExampleEDF} does not depend upon corrections to the instrument geometric factor or anode efficiencies.  It only requires knowledge of the energy bin values, which is one of the most well known parameters in a particle detector.  The ability to directly determine $\phi{\scriptstyle_{sc}}$ from the particle data enables missions that do not have a direct measure of $\phi{\scriptstyle_{sc}}$ from an electric field instrument to properly calculate velocity moments and/or perform analysis on the electron velocity distributions.  It can also provide a sanity check to help calibrate electric field measurements that directly measure $\phi{\scriptstyle_{sc}}$.

\indent  As previous efforts have shown \citep[e.g.,][]{genot04a, lavraud16a, song97a}, properly correcting the electron measurements for $\phi{\scriptstyle_{sc}}$ is critical for calculating any velocity moments and/or instability analysis.  In fact, poor estimates of $\phi{\scriptstyle_{sc}}$ or low resolution instrumentation or instruments with $E{\scriptstyle_{min}}$ $>$ $\phi{\scriptstyle_{sc}}$ can result in velocity moment errors from $\sim$10--30\% to over 100\% \citep[e.g.,][]{song97a, wilsoniii22h}.  Therefore, it is essential that electron thermal particle detectors be designed such that $E{\scriptstyle_{min}}$ $<$ $\phi{\scriptstyle_{sc}}$.

%%----------------------------------------------------------------------------------------
%%  Acknowledgements
%%----------------------------------------------------------------------------------------
\acknowledgments
\noindent  Analysis software \citep[][]{wilsoniii21c} used herein can be found at: \\
\noindent  \url{https://github.com/lynnbwilsoniii/wind\_3dp\_pros}.  \\
\noindent  The \emph{Wind} 3DP level zero data files are publicly available at: \\
\noindent  \url{http://sprg.ssl.berkeley.edu/wind3dp/data/wi/3dp/lz/}.  \\
\noindent  Some of the results presented in this document rely on data measured from the Thermosphere Ionosphere Mesosphere Energetics and Dynamics (TIMED) Solar EUV Experiment (SEE). These data are available from the TIMED SEE website at \\
\noindent  \url{https://lasp.colorado.edu/see/data/}.  \\
\noindent  These data were accessed via the LASP Interactive Solar Irradiance Datacenter (LISIRD) at: \\
\noindent  \url{https://lasp.colorado.edu/lisird/}. \\
\noindent  The resulting dataset of \emph{Wind} spacecraft floating potential measurements are found at \citet[][]{wilsoniii23h}.  Work at the University of California Berkeley was supported by NASA grant 80NSSC20K0708 and NSF grant 2203319.

%\clearpage
\appendix
%%@@@@@@@@@@@@@@@@@@@@@@@@@@@@@@@@@@@@@@@@@@@@@@@@@@@@@@@@@@@@@@@@@@@@@@@@@@@@@@@@@@@@@@@@
%%  Appendix:  Definitions
%%@@@@@@@@@@@@@@@@@@@@@@@@@@@@@@@@@@@@@@@@@@@@@@@@@@@@@@@@@@@@@@@@@@@@@@@@@@@@@@@@@@@@@@@@
\phantomsection   %%  Fix reference link
\section{Definitions}  \label{app:Definitions}

\indent  In this appendix the symbols and notation used throughout will be defined.  All direction-dependent parameters we use the subscript $j$ to represent the direction where $j$ $=$ $tot$ for the entire distribution, $j$ $=$ $\parallel$ for the the parallel direction, and $j$ $=$ $\perp$ for the perpendicular direction, where parallel/perpendicular is with respect to the quasi-static magnetic field vector, $\mathbf{B}{\scriptstyle_{o}}$ [nT].   Below are the symbol/parameters definitions:

\begin{itemize}[itemsep=0pt,parsep=0pt,topsep=0pt]
  \item[]  \textit{one-variable statistics}
  \begin{itemize}[itemsep=0pt,parsep=0pt,topsep=0pt]
    \item  $X{\scriptstyle_{min}}$ $\equiv$ minimum
    \item  $X{\scriptstyle_{max}}$ $\equiv$ maximum
    \item  $\bar{X}$ $\equiv$ mean
    \item  $\tilde{X}$ $\equiv$ median
    \item  $X{\scriptstyle_{25\%}}$ $\equiv$ lower quartile
    \item  $X{\scriptstyle_{75\%}}$ $\equiv$ upper quartile
    \item  $X{\scriptstyle_{y\%}}$ $\equiv$ y$^{th}$ percentile
    \item  $\sigma$ $\equiv$ standard deviation
    \item  $\sigma^{2}$ $\equiv$ variance
  \end{itemize}
  \item[]  \textit{fundamental parameters}
  \begin{itemize}[itemsep=0pt,parsep=0pt,topsep=0pt]
    \item  $\varepsilon{\scriptstyle_{o}}$ $\equiv$ permittivity of free space
    \item  $\mu{\scriptstyle_{o}}$ $\equiv$ permeability of free space
    \item  $c$ $\equiv$ speed of light in vacuum [$km \ s^{-1}$] $=$ $\left( \varepsilon{\scriptstyle_{o}} \ \mu{\scriptstyle_{o}} \right)^{-1/2}$
    \item  $k{\scriptstyle_{B}}$ $\equiv$ the Boltzmann constant [$J \ K^{-1}$]
    \item  $e$ $\equiv$ the fundamental charge [$C$]
  \end{itemize}
  \item[]  \textit{plasma parameters}
  \begin{itemize}[itemsep=0pt,parsep=0pt,topsep=0pt]
    \item  $n{\scriptstyle_{s}}$ $\equiv$ the number density [$cm^{-3}$] of species $s$
    \item  $m{\scriptstyle_{s}}$ $\equiv$ the mass [$kg$] of species $s$
    \item  $Z{\scriptstyle_{s}}$ $\equiv$ the charge state of species $s$
    \item  $q{\scriptstyle_{s}}$ $\equiv$ the charge [$C$] of species $s$ $=$ $Z{\scriptstyle_{s}} \ e$
    \item  $T{\scriptstyle_{s, j}}$ $\equiv$ the scalar temperature [$eV$] of the j$^{th}$ component of species $s$
    \item  $V{\scriptstyle_{Ts, j}}$ $=$ $\sqrt{ \tfrac{ 2 \ k{\scriptstyle_{B}} \ T{\scriptstyle_{s, j}} }{ m{\scriptstyle_{s}} } }$ $\equiv$ the most probable thermal speed [$km \ s^{-1}$] of a one-dimensional velocity distribution
    \item  $\mathbf{V}{\scriptstyle_{os}}$ $\equiv$ the drift velocity [$km \ s^{-1}$] of species $s$ in the plasma relative to a specified rest frame
    \item  $\phi{\scriptstyle_{sc}}$ $\equiv$ the scalar, quasi-static spacecraft potential [eV] \citep[e.g.,][]{pulupa14a, scime94a}
    \item  $E{\scriptstyle_{min}}$ $\equiv$ the minimum energy bin midpoint value [eV] of a particle detector
    \item  $E{\scriptstyle_{max}}$ $\equiv$ the maximum energy allowed when finding $\phi{\scriptstyle_{sc}}$ from the particle data
    \item  $j{\scriptstyle_{th,s}}$ $=$ $q{\scriptstyle_{s}} \ n{\scriptstyle_{s}} \ V{\scriptstyle_{Ts, tot}}$ $\equiv$ the thermal current density [$\mu A \ m^{-2}$] due to the ambient plasma species $s$
    \item  $j{\scriptstyle_{o,s}}$ $=$ $q{\scriptstyle_{s}} \ n{\scriptstyle_{s}} \ \lvert \mathbf{V}{\scriptstyle_{os}} \rvert$ $\equiv$ current density [$\mu A \ m^{-2}$] due to ambient plasma species $s$ drifting relative to spacecraft
    \item  $j{\scriptstyle_{ph}}$ $\equiv$ the photoelectron current density [$\mu A \ m^{-2}$]
    \item  $j{\scriptstyle_{2nd,s}}$ $\equiv$ the current density [$\mu A \ m^{-2}$] due to secondary particles of species $s$ emitted from the spacecraft due to impact ionization
    \item  $f{\scriptstyle_{s}}\left( E \right)$ $\equiv$ energy distribution function (EDF) of particle species $s$
    \item  $f{\scriptstyle_{med}}\left( E \right)$ $\equiv$ the median of $f{\scriptstyle_{s}}\left( E \right)$ at each energy over all solid angles
  \end{itemize}
\end{itemize}

\clearpage
%%@@@@@@@@@@@@@@@@@@@@@@@@@@@@@@@@@@@@@@@@@@@@@@@@@@@@@@@@@@@@@@@@@@@@@@@@@@@@@@@@@@@@@@@@
%%  Appendix:  Extra Statistics
%%@@@@@@@@@@@@@@@@@@@@@@@@@@@@@@@@@@@@@@@@@@@@@@@@@@@@@@@@@@@@@@@@@@@@@@@@@@@@@@@@@@@@@@@@
\phantomsection   %%  Fix reference link
\section{Extra Statistics}  \label{app:ExtraStatistics}

\indent  In this appendix we provide additional one-variable statistics of our estimates for $\phi{\scriptstyle_{sc}}$ separated by EDF types.  Note that we only show Types A and B here, as Types C1 and C2 have $E{\scriptstyle_{min}}$ $>$ $\phi{\scriptstyle_{sc}}$ and thus, we cannot properly measure $\phi{\scriptstyle_{sc}}$ from the thermal electron EDFs.

%%========================================================================================
%%  Table:  Type A EDFs
%%========================================================================================
\startlongtable  %%  In case table runs long
\begin{deluxetable}{| l | c | c | c | c | c | c | c | c |}
  \tabletypesize{\normalsize}    %%  Table font size
  %%  Table Caption
  \tablecaption{Type A EDFs \label{tab:TypeAEDFs}}
  %%  Column Headers
  \tablehead{\colhead{$\phi{\scriptstyle_{sc}}$ [eV]} & \colhead{$X{\scriptstyle_{min}}$} & \colhead{$X{\scriptstyle_{5\%}}$} & \colhead{$X{\scriptstyle_{25\%}}$} & \colhead{$\bar{X}$} & \colhead{$\tilde{X}$} & \colhead{$X{\scriptstyle_{75\%}}$} & \colhead{$X{\scriptstyle_{95\%}}$} & \colhead{$X{\scriptstyle_{max}}$}}
  %%======================================================================================
  %%  Table Body
  %%======================================================================================
  \startdata
  \multicolumn{9}{ |c| }{\textit{All \totalnvdfsAs~Type A EDFs, Lower Bound}} \\
  \hline
  %%                    Min    5%    25%    Avg    Med    75%    95%     Max
  \textbf{Method 1}  & 1.00 & 1.00 & 5.18 & 4.96 & 5.18 & 5.18 & 5.18 & 17.3  \\
  \textbf{Method 2}  & 1.00 & 1.00 & 5.67 & 6.54 & 7.09 & 7.09 & 8.48 & 17.3  \\
  \textbf{Method 3}  & 1.00 & 1.00 & 5.18 & 5.14 & 5.18 & 5.93 & 5.93 & 12.7  \\
  \textbf{Method 4}  & 4.91 & 5.13 & 5.93 & 7.02 & 7.41 & 7.41 & 9.27 & 29.7  \\
  \hline
  \multicolumn{9}{ |c| }{\textit{All \totalnvdfsAs~Type A EDFs, Upper Bound}} \\
  \hline
  %%                    Min    5%    25%    Avg    Med    75%    95%     Max
  \textbf{Method 1}  & 4.91 & 5.13 & 29.7 & 27.7 & 29.7 & 29.7 & 29.7 & 29.7  \\
  \textbf{Method 2}  & 4.91 & 5.13 & 6.20 & 7.37 & 7.75 & 7.75 & 9.27 & 19.0  \\
  \textbf{Method 3}  & 4.91 & 5.13 & 12.7 & 13.2 & 12.7 & 12.7 & 18.1 & 29.7  \\
  \textbf{Method 4}  & 4.91 & 5.13 & 7.09 & 7.28 & 7.41 & 7.41 & 9.27 & 29.7  \\
  \hline
  \enddata
  %%======================================================================================
  %%  Table Notes
  %%======================================================================================
  \tablecomments{Symbol definitions are the same as in Table \ref{tab:AllEDFs}.}
\vspace{-20pt}
\end{deluxetable}
%%========================================================================================
%%  Table:  Type A EDFs
%%========================================================================================

\onecolumngrid  %%  Need this because deluxetable seems to disable single-column format
\indent  Table \ref{tab:TypeAEDFs} shows one-variable statistics for only Type A EDFs while Table \ref{tab:TypeBEDFs} shows one-variable statistics for only Type B EDFs.  Note that we have added the 25$^{th}$ and 75$^{th}$ percentiles to these tables for extra detail.

%%========================================================================================
%%  Table:  Type B EDFs
%%========================================================================================
\startlongtable  %%  In case table runs long
\begin{deluxetable}{| l | c | c | c | c | c | c | c | c |}
  \tabletypesize{\normalsize}    %%  Table font size
  %%  Table Caption
  \tablecaption{Type B EDFs \label{tab:TypeBEDFs}}
  %%  Column Headers
  \tablehead{\colhead{$\phi{\scriptstyle_{sc}}$ [eV]} & \colhead{$X{\scriptstyle_{min}}$} & \colhead{$X{\scriptstyle_{5\%}}$} & \colhead{$X{\scriptstyle_{25\%}}$} & \colhead{$\bar{X}$} & \colhead{$\tilde{X}$} & \colhead{$X{\scriptstyle_{75\%}}$} & \colhead{$X{\scriptstyle_{95\%}}$} & \colhead{$X{\scriptstyle_{max}}$}}
  %%======================================================================================
  %%  Table Body
  %%======================================================================================
  \startdata
  \multicolumn{9}{ |c| }{\textit{All \totalnvdfsBs~Type B EDFs, Lower Bound}} \\
  \hline
  %%                    Min    5%    25%    Avg    Med    75%    95%     Max
  \textbf{Method 1}  & 1.00 & 5.18 & 5.18 & 5.18 & 5.18 & 5.18 & 5.18 & 6.48  \\
  \textbf{Method 2}  & 1.00 & 8.48 & 8.87 & 9.96 & 9.27 & 11.6 & 12.7 & 27.1  \\
  \textbf{Method 3}  & 1.00 & 5.18 & 5.18 & 5.31 & 5.18 & 5.18 & 5.93 & 16.6  \\
  \textbf{Method 4}  & 5.13 & 5.93 & 6.48 & 9.26 & 9.27 & 9.70 & 18.1 & 29.7  \\
  \hline
  \multicolumn{9}{ |c| }{\textit{All \totalnvdfsBs~Type B EDFs, Upper Bound}} \\
  \hline
  %%                    Min    5%    25%    Avg    Med    75%    95%     Max
  \textbf{Method 1}  & 5.13 & 29.7 & 29.7 & 29.3 & 29.7 & 29.7 & 29.7 & 29.7  \\
  \textbf{Method 2}  & 5.13 & 9.27 & 9.70 & 10.9 & 10.1 & 12.7 & 13.9 & 29.7  \\
  \textbf{Method 3}  & 5.13 & 5.93 & 6.48 & 12.5 & 12.7 & 18.1 & 18.1 & 29.7  \\
  \textbf{Method 4}  & 5.13 & 5.93 & 6.48 & 10.7 & 9.70 & 12.7 & 18.1 & 29.7  \\
  \hline
  \enddata
  %%======================================================================================
  %%  Table Notes
  %%======================================================================================
  \tablecomments{Symbol definitions are the same as in Table \ref{tab:AllEDFs}.}
\vspace{-20pt}
\end{deluxetable}
%%========================================================================================
%%  Table:  Type B EDFs
%%========================================================================================

\clearpage
%%@@@@@@@@@@@@@@@@@@@@@@@@@@@@@@@@@@@@@@@@@@@@@@@@@@@@@@@@@@@@@@@@@@@@@@@@@@@@@@@@@@@@@@@@
%%  Appendix:  Public Dataset
%%@@@@@@@@@@@@@@@@@@@@@@@@@@@@@@@@@@@@@@@@@@@@@@@@@@@@@@@@@@@@@@@@@@@@@@@@@@@@@@@@@@@@@@@@
\phantomsection   %%  Fix reference link
\section{Public Dataset}  \label{app:PublicDataset}

\indent  In this section we describe the variables that are included in the public dataset \citep[][]{wilsoniii23h} and list where to find them.

\indent  Below we describe the variables in the public dataset and we use generic characters to define the number of elements in an array.  We define $N$ as the total number of EDFs (e.g., $\sim$\totalnvdfsall~for entire dataset, but will vary by year), $P$ as the number of pitch-angle bins (i.e., 3), and $R$ as the number of bounds or limits for any given estimtae (i.e., 2).  The array dimensions will be shown in brackets, with each dimension separated by a comma.  The public dataset will include the following variables:

\begin{description}[itemsep=0pt,parsep=0pt,topsep=0pt]
  \item[Unix]     [N,R]-Element (numeric) array of Unix\footnote{Note that the times are not really Unix times, as they include leap seconds from UTC time conversions.  The UMN software discussed in the Acknowledgements Section below handles this accordingly.} start and end times [seconds]
  \item[YMDB]     [N,R]-Element (character string) array of UTC start and end times with format `YYYY-MM-DD/hh:mm:ss.xxx'
  \item[PosCurv]  [N,P,R]-Element (numeric) array of $\phi{\scriptstyle_{sc}}$ [eV] estimates calculated from \textbf{Method 1} (i.e., positive curvature region)
  \item[InflPnt]  [N,P,R]-Element (numeric) array of $\phi{\scriptstyle_{sc}}$ [eV] estimates calculated from \textbf{Method 2} (i.e., inflection point of $f{\scriptstyle_{e}}\left( E \right)$)
  \item[MinMaxC]  [N,P,R]-Element (numeric) array of $\phi{\scriptstyle_{sc}}$ [eV] estimates calculated from \textbf{Method 3} (i.e., region of minimum and maximum curvature)
  \item[MinEflx]  [N,P]-Element (numeric) array of $\phi{\scriptstyle_{sc}}$ [eV] estimates calculated from \textbf{Method 4} (i.e., local minimum in $f{\scriptstyle_{e}}\left( E \right)$)
  \item[MinEner]  [N]-Element (numeric) array of the instrument $E{\scriptstyle_{min}}$ values [eV]
  \item[MaxEner]  [N]-Element (numeric) array of the $E{\scriptstyle_{max}}$ values [eV] chosen a priori to bound the solutions for $\phi{\scriptstyle_{sc}}$
  \item[EDFType]  [N]-Element (character string) array of EDF types (e.g., `A' for Type A)
  \item[QFlag]    [N]-Element (integer) array of quality flags based upon EDF types (e.g., 4 for Type A)
\end{description}

\indent  Note that the values for each of these have not been altered from those returned by the automated software performing these tests.  For instance, we have not adjusted results from \textbf{Method 3} for EDFs of Type C1 or C2.  We left the original values in place and allow the user to alter the upper/lower bounds as they see fit for their own purposes.  In general, for EDFs of Type C1 or C2 with shapes matching that shown in Figure \ref{fig:FourShapes}, $\phi{\scriptstyle_{sc}}$ $<$ $E{\scriptstyle_{min}}$ for all four methods.  This is why they are given a QF of zero.  However, one should note that such EDFs and associated $\phi{\scriptstyle_{sc}}$ are not useless or meaningless.  These times are intervals where we can confidently define an absolute upper bound on $\phi{\scriptstyle_{sc}}$.  They also correspond to times when $\phi{\scriptstyle_{sc}}$ is very small, relative to the typical values shown in Figure \ref{fig:2DHist} and Table \ref{tab:AllEDFs}.

\indent  The dataset will consist of a list of yearly ASCII files from January 1, 2005 to January 1, 2022.  More years can be added as time progresses to expand the dataset.  This dataset is the first step towards generating a properly calibrated electron velocity moment data product for public consumption.  The next step requires the knowledge of the total electron density from the upper hybrid line, a dataset that is currently in production \citep[e.g., see][for details]{martinovic20b}.  The combination of accurate estimates from these two quantities will allow us to update the anode calibrations\footnote{In the unit conversion process, this calibration enters through a modification to the optical geometric factor for each energy-angle bin.  It is important to note that these calibrations do not affect the detector energy bin values, i.e., they will not affect $\phi{\scriptstyle_{sc}}$ for the methods used herein.} for the \emph{Wind} 3DP thermal electron detector, EESA Low.  Note that spacecraft charging can also affect the trajectories of incident low energy particles \citep[e.g., see][for details]{barrie19a}.  However, this would require an accurate model of the \emph{Wind} spacecraft surface materials and the evolution of their work functions over time, which is beyond the scope of this effort.

\indent  Finally, we should note that the dataset \citet[][]{wind3dpechsfits23a} provided by and discussed in \citet[][]{salem23a} does not overlap with this new dataset and it spans the interval when $E{\scriptstyle_{min}}$ $>$ $\phi{\scriptstyle_{sc}}$, early in the \emph{Wind} mission.  Thus, \citet[][]{salem23a} relied on different techniques to estimate $\phi{\scriptstyle_{sc}}$ than the methods described herein.  The measurements in \citet[][]{salem23a} are also subject to smaller variations in the anode calibrations than the measurements made later in the mission (i.e., the data herein).  This was shown in a recent study \citep[e.g., see appendices of][]{wilsoniii19a} in some example anode correction calculations for the time period examined by \citet[][]{salem23a}.  Thus, they were able to estimate $\phi{\scriptstyle_{sc}}$ numerically from the difference between the total electron density calculated from the upper hybrid line and that calculated by numerically integrating the velocity distribution.  They iteratively modify $\phi{\scriptstyle_{sc}}$ until the numerically integrated electron density matches that from the upper hybrid line.  This is a well known and valid approach when $E{\scriptstyle_{min}}$ $>$ $\phi{\scriptstyle_{sc}}$, assuming the instrument anodes and energy-angle geometric factors are properly calibrated.

%\clearpage
%%----------------------------------------------------------------------------------------
%%  Bibliography
%%----------------------------------------------------------------------------------------
%\bibliography{/Users/lbwilson/Desktop/Lynn_B_Wilson_III/LaTeX/Bibliographies/my_bib_maker}

\begin{thebibliography}{}
\expandafter\ifx\csname natexlab\endcsname\relax\def\natexlab#1{#1}\fi
\providecommand{\url}[1]{\href{#1}{#1}}
\providecommand{\dodoi}[1]{doi:~\href{http://doi.org/#1}{\nolinkurl{#1}}}
\providecommand{\doeprint}[1]{\href{http://ascl.net/#1}{\nolinkurl{http://ascl.net/#1}}}
\providecommand{\doarXiv}[1]{\href{https://arxiv.org/abs/#1}{\nolinkurl{https://arxiv.org/abs/#1}}}

\bibitem[{{Bame} {et~al.}(1968){Bame}, {Hundhausen}, {Asbridge}, \&
  {Strong}}]{bame68a}
{Bame}, S.~J., {Hundhausen}, A.~J., {Asbridge}, J.~R., \& {Strong}, I.~B. 1968,
  Phys. Rev. Lett., 20, 393, \dodoi{10.1103/PhysRevLett.20.393}

\bibitem[{{Barrie} {et~al.}(2019){Barrie}, {Cipriani}, {Escoubet},
  {Toledo-Redondo}, {Nakamura}, {Torkar}, {Sternovsky}, {Elkington},
  {Gershman}, {Giles}, \& {Schiff}}]{barrie19a}
{Barrie}, A.~C., {Cipriani}, F., {Escoubet}, C.~P., {et~al.} 2019, Phys.
  Plasmas, 26, 103504, \dodoi{10.1063/1.5119344}

\bibitem[{{Besse} \& {Rubin}(1980)}]{besse80a}
{Besse}, A.~L., \& {Rubin}, A.~G. 1980, J. Geophys. Res., 85, 2324,
  \dodoi{10.1029/JA085iA05p02324}

\bibitem[{{Bochsler} {et~al.}(1985){Bochsler}, {Geis}, \& {Joos}}]{bochsler85a}
{Bochsler}, P., {Geis}, J., \& {Joos}, R. 1985, J. Geophys. Res., 90, 10779,
  \dodoi{10.1029/JA090iA11p10779}

\bibitem[{{Bougeret} {et~al.}(1995){Bougeret}, {Kaiser}, {Kellogg}, {Manning},
  {Goetz}, {Monson}, {Monge}, {Friel}, {Meetre}, {Perche}, {Sitruk}, \&
  {Hoang}}]{bougeret95a}
{Bougeret}, J.-L., {Kaiser}, M.~L., {Kellogg}, P.~J., {et~al.} 1995, Space Sci.
  Rev., 71, 231, \dodoi{10.1007/BF00751331}

\bibitem[{{Ergun} {et~al.}(2016){Ergun}, {Tucker}, {Westfall}, {Goodrich},
  {Malaspina}, {Summers}, {Wallace}, {Karlsson}, {Mack}, {Brennan}, {Pyke},
  {Withnell}, {Torbert}, {Macri}, {Rau}, {Dors}, {Needell}, {Lindqvist},
  {Olsson}, \& {Cully}}]{ergun16a}
{Ergun}, R.~E., {Tucker}, S., {Westfall}, J., {et~al.} 2016, Space Sci. Rev.,
  199, 167, \dodoi{10.1007/s11214-014-0115-x}

\bibitem[{{Garrett}(1981)}]{garrett81a}
{Garrett}, H.~B. 1981, Rev. Geophys. Space Phys., 19, 577

\bibitem[{{Geach} {et~al.}(2005){Geach}, {Schwartz}, {G{\'e}not}, {Moullard},
  {Lahiff}, \& {Fazakerley}}]{geach05a}
{Geach}, J., {Schwartz}, S.~J., {G{\'e}not}, V., {et~al.} 2005, Ann. Geophys.,
  23, 931, \dodoi{10.5194/angeo-23-931-2005}

\bibitem[{{G{\'e}not} \& {Schwartz}(2004)}]{genot04a}
{G{\'e}not}, V., \& {Schwartz}, S. 2004, Ann. Geophys., 22, 2073,
  \dodoi{10.5194/angeo-22-2073-2004}

\bibitem[{{Gershman} {et~al.}(2017){Gershman}, {Avanov}, {Boardsen}, {Dorelli},
  {Gliese}, {Barrie}, {Schiff}, {Paterson}, {Torbert}, {Giles}, \&
  {Pollock}}]{gershman17a}
{Gershman}, D.~J., {Avanov}, L.~A., {Boardsen}, S.~A., {et~al.} 2017, J.
  Geophys. Res., 122, 11548, \dodoi{10.1002/2017JA024518}

\bibitem[{{Gloeckler} {et~al.}(1998){Gloeckler}, {Cain}, {Ipavich}, {Tums},
  {Bedini}, {Fisk}, {Zurbuchen}, {Bochsler}, {Fischer}, {Wimmer-Schweingruber},
  {Geiss}, \& {Kallenbach}}]{gloeckler98a}
{Gloeckler}, G., {Cain}, J., {Ipavich}, F.~M., {et~al.} 1998, Space Sci. Rev.,
  86, 497, \dodoi{10.1023/A:1005036131689}

\bibitem[{{Grard} {et~al.}(1983){Grard}, {Knott}, \& {Pedersen}}]{grard83a}
{Grard}, R., {Knott}, K., \& {Pedersen}, A. 1983, Space Sci. Rev., 34, 289,
  \dodoi{10.1007/BF00175284}

\bibitem[{{Grard}(1973)}]{grard73a}
{Grard}, R.~J.~L. 1973, J. Geophys. Res., 78, 2885,
  \dodoi{10.1029/JA078i016p02885}

\bibitem[{{Halekas} {et~al.}(2020){Halekas}, {Whittlesey}, {Larson},
  {McGinnis}, {Maksimovic}, {Berthomier}, {Kasper}, {Case}, {Korreck},
  {Stevens}, {Klein}, {Bale}, {MacDowall}, {Pulupa}, {Malaspina}, {Goetz}, \&
  {Harvey}}]{halekas20a}
{Halekas}, J.~S., {Whittlesey}, P., {Larson}, D.~E., {et~al.} 2020, Astrophys.
  J. Suppl., 246, 22, \dodoi{10.3847/1538-4365/ab4cec}

\bibitem[{{Jian} {et~al.}(2006{\natexlab{a}}){Jian}, {Russell}, {Luhmann}, \&
  {Skoug}}]{jian06b}
{Jian}, L., {Russell}, C.~T., {Luhmann}, J.~G., \& {Skoug}, R.~M.
  2006{\natexlab{a}}, Solar Phys., 239, 393, \dodoi{10.1007/s11207-006-0133-2}

\bibitem[{{Jian} {et~al.}(2006{\natexlab{b}}){Jian}, {Russell}, {Luhmann}, \&
  {Skoug}}]{jian06a}
---. 2006{\natexlab{b}}, Solar Phys., 239, 337,
  \dodoi{10.1007/s11207-006-0132-3}

\bibitem[{{Lai} {et~al.}(2017){Lai}, {Martinez-Sanchez}, {Cahoy}, {Thomsen},
  {Shprits}, {Lohmeyer}, \& {Wong}}]{lai17a}
{Lai}, S.~T., {Martinez-Sanchez}, M., {Cahoy}, K., {et~al.} 2017, IEEE Trans.
  Plasma Sci., 45, 2875, \dodoi{10.1109/TPS.2017.2751002}

\bibitem[{{Lavraud} \& {Larson}(2016)}]{lavraud16a}
{Lavraud}, B., \& {Larson}, D.~E. 2016, J. Geophys. Res., 121, 8462,
  \dodoi{10.1002/2016JA022591}

\bibitem[{{Lepri} {et~al.}(2013){Lepri}, {Landi}, \& {Zurbuchen}}]{lepri13a}
{Lepri}, S.~T., {Landi}, E., \& {Zurbuchen}, T.~H. 2013, Astrophys. J., 768,
  94, \dodoi{10.1088/0004-637X/768/1/94}

\bibitem[{{Lin} {et~al.}(1995){Lin}, {Anderson}, {Ashford}, {Carlson},
  {Curtis}, {Ergun}, {Larson}, {McFadden}, {McCarthy}, {Parks}, {R\`{e}me},
  {Bosqued}, {Coutelier}, {Cotin}, {D'Uston}, {Wenzel}, {Sanderson}, {Henrion},
  {Ronnet}, \& {Paschmann}}]{lin95a}
{Lin}, R.~P., {Anderson}, K.~A., {Ashford}, S., {et~al.} 1995, Space Sci. Rev.,
  71, 125, \dodoi{10.1007/BF00751328}

\bibitem[{{Lindqvist} {et~al.}(2016){Lindqvist}, {Olsson}, {Torbert}, {King},
  {Granoff}, {Rau}, {Needell}, {Turco}, {Dors}, {Beckman}, {Macri}, {Frost},
  {Salwen}, {Eriksson}, {{\AA}hl{\'e}n}, {Khotyaintsev}, {Porter},
  {Lappalainen}, {Ergun}, {Wermeer}, \& {Tucker}}]{lindqvist16a}
{Lindqvist}, P.-A., {Olsson}, G., {Torbert}, R.~B., {et~al.} 2016, Space Sci.
  Rev., 199, 137, \dodoi{10.1007/s11214-014-0116-9}

\bibitem[{{Martinovi{\'c}} {et~al.}(2020){Martinovi{\'c}}, {Klein}, {Gramze},
  {Jain}, {Maksimovi{\'c}}, {Zaslavsky}, {Salem}, {Zouganelis}, \&
  {Simi{\'c}}}]{martinovic20b}
{Martinovi{\'c}}, M.~M., {Klein}, K.~G., {Gramze}, S.~R., {et~al.} 2020, J.
  Geophys. Res., 125, e28113, \dodoi{10.1029/2020JA028113}

\bibitem[{{Maruca} {et~al.}(2013){Maruca}, {Bale}, {Sorriso-Valvo}, {Kasper},
  \& {Stevens}}]{maruca13b}
{Maruca}, B.~A., {Bale}, S.~D., {Sorriso-Valvo}, L., {Kasper}, J.~C., \&
  {Stevens}, M.~L. 2013, Phys. Rev. Lett., 111, 241101,
  \dodoi{10.1103/PhysRevLett.111.241101}

\bibitem[{{Meyer-Vernet} \& {Perche}(1989)}]{meyervernet89a}
{Meyer-Vernet}, N., \& {Perche}, C. 1989, J. Geophys. Res., 94, 2405,
  \dodoi{10.1029/JA094iA03p02405}

\bibitem[{{Pedersen}(1995)}]{pedersen95a}
{Pedersen}, A. 1995, Ann. Geophys., 13, 118, \dodoi{10.1007/s00585-995-0118-8}

\bibitem[{{Phillips} {et~al.}(1993){Phillips}, {Bame}, {Gosling}, {McComas},
  {Goldstein}, \& {Balogh}}]{phillips93a}
{Phillips}, J.~L., {Bame}, S.~J., {Gosling}, J.~T., {et~al.} 1993, Adv. Space
  Res., 13, 47, \dodoi{10.1016/0273-1177(93)90389-S}

\bibitem[{{Pulupa} {et~al.}(2014){Pulupa}, {Bale}, {Salem}, \&
  {Horaites}}]{pulupa14a}
{Pulupa}, M.~P., {Bale}, S.~D., {Salem}, C., \& {Horaites}, K. 2014, J.
  Geophys. Res., 119, 647, \dodoi{10.1002/2013JA019359}

\bibitem[{{Richardson} {et~al.}(2018){Richardson}, {Mays}, \&
  {Thompson}}]{richardson18a}
{Richardson}, I.~G., {Mays}, M.~L., \& {Thompson}, B.~J. 2018, Space Weather,
  16, 1862, \dodoi{10.1029/2018SW002032}

\bibitem[{{Salem} {et~al.}(2001){Salem}, {Bosqued}, {Larson}, {Mangeney},
  {Maksimovic}, {Perche}, {Lin}, \& {Bougeret}}]{salem01a}
{Salem}, C., {Bosqued}, J.-M., {Larson}, D.~E., {et~al.} 2001, J. Geophys.
  Res., 106, 21701, \dodoi{10.1029/2001JA900031}

\bibitem[{{Salem} {et~al.}(2003){Salem}, {Hubert}, {Lacombe}, {Bale},
  {Mangeney}, {Larson}, \& {Lin}}]{salem03a}
{Salem}, C., {Hubert}, D., {Lacombe}, C., {et~al.} 2003, Astrophys. J., 585,
  1147, \dodoi{10.1086/346185}

\bibitem[{{Salem} {et~al.}(2023){Salem}, {Pulupa}, {Bale}, \&
  {Verscharen}}]{salem23a}
{Salem}, C.~S., {Pulupa}, M., {Bale}, S.~D., \& {Verscharen}, D. 2023, Astron.
  \& Astrophys., 675, A162, \dodoi{10.1051/0004-6361/202141816}

\bibitem[{{Salem} \& {Pulupa}(2023)}]{wind3dpechsfits23a}
{Salem}, C.~S., \& {Pulupa}, M.~P. 2023, \emph{Wind} 3DP Electron Core Halo
  Strahl Moments, 96s, year files, 2.5.0,  NASA Space Physics Data Facility,
  \dodoi{10.48322/rgf7-3h67}.
\newblock \url{https://doi.org/10.48322/rgf7-3h67}

\bibitem[{{Scime} {et~al.}(1994){Scime}, {Phillips}, \& {Bame}}]{scime94a}
{Scime}, E.~E., {Phillips}, J.~L., \& {Bame}, S.~J. 1994, J. Geophys. Res., 99,
  14769, \dodoi{10.1029/94JA00489}

\bibitem[{{Scudder} {et~al.}(2000){Scudder}, {Cao}, \& {Mozer}}]{scudder00a}
{Scudder}, J.~D., {Cao}, X., \& {Mozer}, F.~S. 2000, J. Geophys. Res., 105,
  21281, \dodoi{10.1029/1999JA900423}

\bibitem[{{Song} {et~al.}(1997){Song}, {Zhang}, \& {Paschmann}}]{song97a}
{Song}, P., {Zhang}, X.~X., \& {Paschmann}, G. 1997, Planet. Space Sci., 45,
  255, \dodoi{10.1016/S0032-0633(96)00087-6}

\bibitem[{{Whipple}(1981)}]{whipple81a}
{Whipple}, E.~C. 1981, Rep. Progr. Phys., 44, 1197,
  \dodoi{10.1088/0034-4885/44/11/002}

\bibitem[{{Wilson III}(2021)}]{wilsoniii21c}
{Wilson III}, L.~B. 2021, Space Plasma Missions IDL Software Library, 1.0.2,
  Zenodo, \dodoi{10.5281/zenodo.6141586}.
\newblock \url{https://doi.org/10.5281/zenodo.6141586}

\bibitem[{{Wilson III} {et~al.}(2022){Wilson III}, {Goodrich}, {Turner},
  {Cohen}, {Whittlesey}, \& {Schwartz}}]{wilsoniii22h}
{Wilson III}, L.~B., {Goodrich}, K.~A., {Turner}, D.~L., {et~al.} 2022, Front.
  Astron. Space Sci., 9, 1063841, \dodoi{10.3389/fspas.2022.1063841}

\bibitem[{{Wilson III} {et~al.}(2023){Wilson III}, {Salem}, \&
  {Bonnell}}]{wilsoniii23h}
{Wilson III}, L.~B., {Salem}, C.~S., \& {Bonnell}, J.~W. 2023, \emph{Wind}
  spacecraft floating potential measurements, 1.0,  Zenodo,
  \dodoi{10.5281/zenodo.8364797}.
\newblock \url{https://doi.org/10.5281/zenodo.8364797}

\bibitem[{{Wilson III} {et~al.}(2018){Wilson III}, {Stevens}, {Kasper},
  {Klein}, {Maruca}, {Bale}, {Bowen}, {Pulupa}, \& {Salem}}]{wilsoniii18b}
{Wilson III}, L.~B., {Stevens}, M.~L., {Kasper}, J.~C., {et~al.} 2018,
  Astrophys. J. Suppl., 236, 41, \dodoi{10.3847/1538-4365/aab71c}

\bibitem[{{Wilson III} {et~al.}(2019{\natexlab{a}}){Wilson III}, {Chen},
  {Wang}, {Schwartz}, {Turner}, {Stevens}, {Kasper}, {Osmane}, {Caprioli},
  {Bale}, {Pulupa}, {Salem}, \& {Goodrich}}]{wilsoniii19a}
{Wilson III}, L.~B., {Chen}, L.-J., {Wang}, S., {et~al.} 2019{\natexlab{a}},
  Astrophys. J. Suppl., 243, \dodoi{10.3847/1538-4365/ab22bd}

\bibitem[{{Wilson III} {et~al.}(2019{\natexlab{b}}){Wilson III}, {Chen},
  {Wang}, {Schwartz}, {Turner}, {Stevens}, {Kasper}, {Osmane}, {Caprioli},
  {Bale}, {Pulupa}, {Salem}, \& {Goodrich}}]{wilsoniii19b}
---. 2019{\natexlab{b}}, Astrophys. J. Suppl., 245,
  \dodoi{10.3847/1538-4365/ab5445}

\bibitem[{{Wilson III} {et~al.}(2021){Wilson III}, {Brosius}, {Gopalswamy},
  {Nieves-Chinchilla}, {Szabo}, {Hurley}, {Phan}, {Kasper}, {Lugaz},
  {Richardson}, {Chen}, {Verscharen}, {Wicks}, \& {TenBarge}}]{wilsoniii21b}
{Wilson III}, L.~B., {Brosius}, A.~L., {Gopalswamy}, N., {et~al.} 2021, Rev.
  Geophys., 59, e2020RG000714, \dodoi{10.1029/2020RG000714}

\bibitem[{{Woods} {et~al.}(2005){Woods}, {Eparvier}, {Bailey}, {Chamberlin},
  {Lean}, {Rottman}, {Solomon}, {Tobiska}, \& {Woodraska}}]{woods05a}
{Woods}, T.~N., {Eparvier}, F.~G., {Bailey}, S.~M., {et~al.} 2005, J. Geophys.
  Res., 110, A01312, \dodoi{10.1029/2004JA010765}

\end{thebibliography}

\end{document}